\documentclass[12pt, onecolumn]{IEEEtran}
\usepackage{amsmath, amsthm, amssymb} 
\usepackage{graphicx}
\usepackage{epstopdf}
\usepackage{verbatim}
\usepackage[T1]{fontenc}
\usepackage[noadjust]{cite}
\usepackage{color}
\usepackage{soul}
\allowdisplaybreaks
\ifCLASSINFOpdf

\else
  
\fi

\begin{document}

\title{Hybrid Beamforming for Large Antenna Arrays with Phase Shifter Selection}

\author{\IEEEauthorblockN{Sohail Payami, Mir Ghoraishi, and Mehrdad Dianati}\\
\IEEEauthorblockA{ Institute for Communication Systems (ICS), \\
5G Innovation Centre (former CCSR), University of Surrey, UK\\
Email: $\lbrace$s.payami, m.ghoraishi, and m.dianati$\rbrace$@surrey.ac.uk}}


\maketitle

\section*{Abstract}
This paper proposes an asymptotically optimal hybrid beamforming solution for large antenna arrays by exploiting the properties of the singular vectors of the channel matrix. It is shown that the elements of the channel matrix with Rayleigh fading follow a normal distribution when large antenna arrays are employed. The proposed beamforming algorithm is effective in both sparse and rich propagation environments, and is applicable for both point-to-point and multiuser scenarios. In addition, a closed-form expression and a lower-bound for the achievable rates are derived when analog and digital phase shifters are employed. It is shown that the performance of the hybrid beamformers using phase shifters with more than 2-bits resolution is comparable with analog phase shifting. A novel phase shifter selection scheme that reduces the power consumption at the phase shifter network is proposed when the wireless channel is modeled by Rayleigh fading. Using this selection scheme, the spectral efficiency can be increased as the power consumption in the phase shifter network reduces. Compared to the scenario that all of the phase shifters are in operation, the simulation results indicate that the spectral efficiency increases when up to $50\%$ of phase shifters are turned off.
\begin{IEEEkeywords}
Hybrid beamforming, large MIMO systems, phase shifter selection.

\end{IEEEkeywords}

\IEEEpeerreviewmaketitle
\section{Introduction}
Multiple-input-multiple-output (MIMO) techniques such as beamforming, precoding and combining can significantly improve the reliability of the transmission and increase the achievable data rates in wireless communication systems. As the number of the antenna elements at the transmitter/receiver increases, higher diversity and multiplexing gains are observed and the channel matrix tends to have favorable conditions \cite{MasiveMIMORusek}. Hence, MIMO systems with large number of antennas have attracted a lot of attention. 

Depending on the structure of the antenna array, analog, digital or hybrid beamformers can be implemented. The analog approach cannot provide multiplexing gains as the antenna array is connected to the transceiver by only a single RF chain \cite{Arrayprocessing}. On the other hand, digital beamformers with a dedicated RF chain per antenna element can use all the degrees of freedom of the channel to transmit multiple symbols simultaneously. However, digital beamforming for large antenna arrays is not suitable for practical applications due to the system complexity, cost and power consumption \cite{AlkhatibHeathMagazine}. In order to provide a tradeoff between performance and cost, hybrid beamformers have been proposed where a small number of RF chains are connected to a large number of antennas through a network of phase shifters \cite{MolischStatisticsBased2006,JSDM2013,JSDM2014,PhaseOnlyLiu2014,
SpatiallySparsePrecodingAyach,SohailICC2015,NiDL15,MOlischAntennaselection2005Journal,
HeathMUMIMO,PerAntennaPi2012,FoadTorontoApril2015,FoadTorontoApril2015Digital}. This type of beamformers show a promising performance even with limited channel state information (CSI) \cite{MolischStatisticsBased2006,JSDM2013,JSDM2014,PhaseOnlyLiu2014}. In order to design hybrid beamformers, however, it is required to solve a complex nonconvex optimization problem due to the constant modulus constraint imposed by the phase shifters \cite{AlkhatibHeathMagazine}. In addition, the phase shifters in practical systems have discrete resolution which converts the optimization to a computationally expensive combinatorial problem \cite{SpatiallySparsePrecodingAyach,SohailICC2015}. 

In the hybrid beamforming approach, it has been shown that the baseband precoder and the RF beamformer can be designed either jointly \cite{SpatiallySparsePrecodingAyach,SohailICC2015,NiDL15} or in two stages \cite{MOlischAntennaselection2005Journal,HeathMUMIMO,PerAntennaPi2012,FoadTorontoApril2015,FoadTorontoApril2015Digital}. For a point-to-point system, a joint design approach based on matching pursuit was proposed when the channel is sparse \cite{SpatiallySparsePrecodingAyach,SohailICC2015}. In this method, firstly the singular vectors of the channel should be calculated, and then the hybrid beamformer is derived by solving an optimization problem to minimize the Euclidean distance between the matrices containing the singular vectors and the weights of the hybrid beamformer. Considering that the calculation of the singular vectors is computationally expensive, the second round of computations can cause sever delays in practical systems. In addition, the achievable spectral efficiency based on \cite{SpatiallySparsePrecodingAyach,SohailICC2015} significantly depends on the number of RF chains in the system and multipath components in the channel. A close to optimal performance for both rich and sparse scattering channels was proposed based on approximating the nonconvex optimization with a convex problem and using an iterative algorithm \cite{NiDL15}. The problem associated with such iterative algorithms is that the convergence time depends on the initial conditions. Hence, the processing time to calculate the parameters of the hybrid beamformer can become a prohibitive factor in real-time systems. In the two stage design approach, the RF beamformer is calculated based on the channel matrix. Then, the baseband precoder takes the impact of the channel matrix and the RF beamformer into account. The optimal hybrid beamformer for a single stream transmission was proposed in \cite{MOlischAntennaselection2005Journal}. Then, based on the simulation results it was shown that hybrid and digital beamformers can achieve a similar spectral efficiency when multiple symbols are transmitted. In this case, the optimality of the hybrid beamformer and its performance closed-form expressions were not derived. Another two stage algorithm that can achieve a close to optimal performance was reported in \cite{PerAntennaPi2012,FoadTorontoApril2015,FoadTorontoApril2015Digital} where the RF beamformer was iteratively calculated. In \cite{SpatiallySparsePrecodingAyach,SohailICC2015,NiDL15,MOlischAntennaselection2005Journal,HeathMUMIMO,
PerAntennaPi2012,FoadTorontoApril2015,FoadTorontoApril2015Digital}, it is not possible to derive the closed-form expression of the performance as computer simulations are necessary to evaluate the performance. Furthermore, due to the computational delays associated with the derivation of the hybrid beamforming weights, the algorithms may not be suitable for practical systems depending on the application. In addition, the power consumption in the RF beamformer will be significantly high as each phase shifter requires some power to operate and hybrid beamformers employ a massive number of these components.

In order to address the aforementioned challenges and facilitate the implementation of hybrid beamformers, two main objectives are followed in this paper. Firstly, an asymptotically optimal hybrid beamforming scheme and the closed-form expressions of the spectral efficiency for both point-to-point and multiuser scenarios in rich and sparse scattering channels are derived. Secondly, a novel phase shifter selection scheme is proposed to simultaneously increase the spectral efficiency and reduce the power consumption in the phase shifter network when rich scattering channels are considered. It is assumed that the rich and sparse scattering channels follow Rayleigh fading and geometry based models, perfect CSI is available at the transmitter and the number of the antennas are large. All the proposed schemes and the closed-form expressions in this paper are derived based on the properties of the singular vectors of the channel matrix. Using the basic characteristics of such vectors, an alternative approach to the solution in \cite{MOlischAntennaselection2005Journal} is presented. It is shown that the performance of the digital beamformers is achievable when the number of the RF chains is two times larger than the number of the transmitted symbols. In order to calculate the hybrid beamformer for the Rayleigh fading scenario when the number of the RF chains and symbols are equal, the distribution of the elements of the singular vectors of the large channel matrix are derived which, to the best knowledge of the authors, has not been previously reported. Based on this distribution, the asymptotically optimal hybrid beamforming schemes for both the point-to-point and multiuser scenarios are derived. Additionally, the closed-form expressions of the spectral efficiencies achieved by the proposed hybrid beamformers are calculated. It is shown that in the solution with optimum performance, the phase shifters in the RF beamformer should be set according to the phase of the elements of the singular vectors of the channel matrix when the number of the antennas are large. The advantages of the proposed approach over the stat-of-the-art is its simplicity, low computational delays and asymptotically optimal behavior. When digital phase shifters are used, a simple but effective hybrid beamforming scheme is proposed and its performance lower-bound is derived. Analytical and simulation results demonstrate that the performance of the proposed scheme with phase shifters with more than 2-bits of resolution is similar to the hybrid beamformer with analog phase shifters. Finally, a novel phase shifter selection scheme and the closed-form expression of its performance bound are presented when the channel matrix follows Rayleigh fading. The advantages of this method are two fold as the power consumption in the RF beamformer network can be reduced and the spectral efficiency can be improved at the same time. Simulation results indicate that the spectral efficiency will increase when up to 50\% of the phase shifters are switched off.

This paper is organized as following, the system model and problem statement of the point-to-point system are described in sections \ref{sec:SystemModel} and \ref{sec:Problem}. In section \ref{sec:proposedalgp2p}, the hybrid beamforming scheme with analog and digital phase shifters are proposed and analyzed. The multiuser scenario and phase shifter selection are investigated in sections \ref{sec:MU} and \ref{sec:PSselect}. Finally, the simulation results, conclusion and future works are presented in sections \ref{sec:imulations} and \ref{sec:conclusion}.

\textit{Notations:} The following notation is used throughout this paper: $\mathbb{R}$ and $\mathbb{C}$ are the field of real and complex numbers. \textbf{A} represents a matrix, \textbf{a} and $\textbf{a}^\ast$ are a vector and its conjugate, respectively. $\textbf{a}_m$ is the $m$th column of \textbf{A} and $\textbf{A}_{1:m}$ is a matrix containing the first $m$ columns of \textbf{A}. $A_{mn}$ and $\vert A_{mn}\vert$ denote the $(m,n)$ element of \textbf{A} and its magnitude. diag($\textbf{A}_1,\:\textbf{A}_2,\:...,\: \textbf{A}_K$) is a diagonal matrix with $\textbf{A}_1,\:\textbf{A}_2,\:..., \: \textbf{A}_K$ on its diagonal. $\textbf{A}^{-1}$, det$( \textbf{A} )$, $\Vert \textbf{A}\Vert$, $\textbf{A}^\text{T}$, $\textbf{A}^\text{H}$, trace(\textbf{A}) denote inverse, determinant, Frobenius norm, transpose, Hermitian and trace of \textbf{A}, respectively. $\mathcal{RN}(\textbf{a},\textbf{A})$ and $\mathcal{CN}(\textbf{a},\textbf{A})$ present a random vector of real and complex Gaussian distributed elements with expected value \textbf{a} and covariance matrix \textbf{A}. Finally, $\textbf{0}_{m\times 1}$, $\textbf{1}_{m\times 1}$ and $\textbf{I}_m$ are a vector of $m$ zeros, $m$ ones and an $m \times m$ identity matrix, respectively.

\section{System Model}
\label{sec:SystemModel}
\begin{figure}
\centering
\includegraphics[width=6.5in]{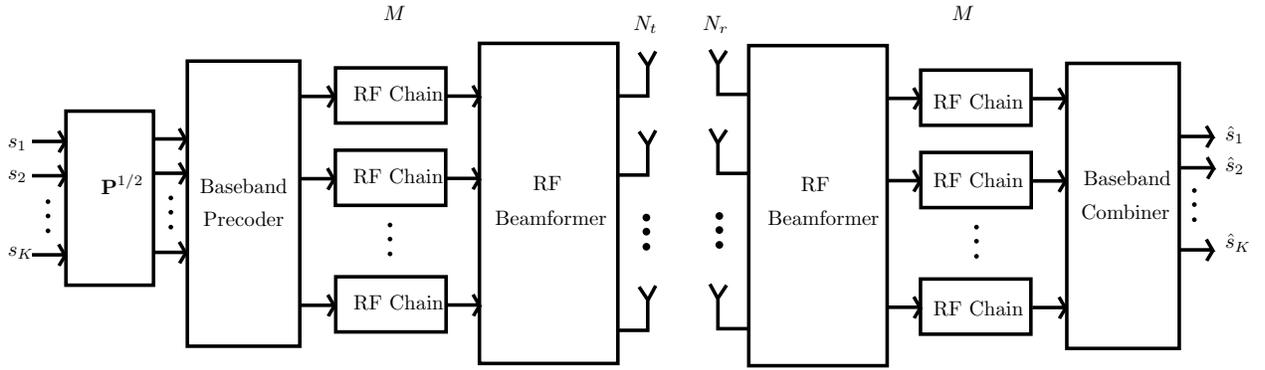}
\caption{The block diagram of a hybrid beamformer for a point-to-point scenario.}
\label{fig:Pointtopoint}
\end{figure}  
In a point-to-point MIMO communication system, the transmitter and the receiver are equipped with $N_\text{t}$ and $N_\text{r}$ antennas, respectively. The transmitter sends a vector $\textbf{s}\in \mathbb{C}^{K \times 1}$ of $K$ symbols to the receiver where E$[\textbf{s}\textbf{s}^\text{H}]=\textbf{I}_K$. The transmit signal vector becomes $\textbf{x}=\sqrt{\dfrac{P_\text{t}}{\Gamma_\text{t}}}\textbf{FP}^{1/2}\textbf{s}$, where $ P_\text{t}$ is the total transmit power, $\textbf{P}\in \mathbb{R}^{K \times K}$ is a diagonal power allocation matrix with $\sum_{k=1}^K P_{kk} \leq 1$, $\textbf{F} \in \mathbb{C}^{N_\text{t} \times K}$ is the precoder matrix and $\Gamma_\text{t}=\text{trace}(\textbf{F}^\text{H}\textbf{F})/K$ is a normalization factor such that $\Vert1/\sqrt{\Gamma_\text{t}} \textbf{F} \Vert^2 = K$. Let $\textbf{H} \in \mathbb{C}^{ N_\text{r} \times N_\text{t}}$ and $\textbf{y} \in \mathbb{C}^{N_\text{r} \times 1}$ denote the normalized channel matrix and the received signal vector. Assuming the noise vector at the receiver antennas $\textbf{z} \sim \mathcal{CN}(\textbf{0}_{N_\text{r} \times 1},\sigma_z^2\textbf{I}_{N_\text{r}})$ has independent and identically distributed (i.i.d.) elements with variance $\sigma_z^2$, the channel input-output relationship is expressed as $\textbf{y}=\textbf{Hx}+\textbf{z}$. Applying the combiner matrix $\textbf{W} \in \mathbb{C}^{ N_\text{r} \times K}$ at the receiver, the input $\hat{\textbf{s}} \in \mathbb{C}^{K \times 1}$ to the detector is
\begin{equation}
\label{eq:GeneralIOPointtoPoint}
\hat{\textbf{s}}=\sqrt{\dfrac{P_\text{t}}{\Gamma_\text{t}\Gamma_\text{r}}}\textbf{W}^\text{H}\textbf{HFP}^{1/2}\textbf{s}+\sqrt{\frac{1}{\Gamma_\text{r}}}\textbf{W}^\text{H}\textbf{z},
\end{equation} 
where $\Gamma_\text{r}=\text{trace}(\textbf{W}^\text{H}\textbf{W})/K$ is a normalization factor such that $\Vert1/\sqrt{\Gamma_\text{r}} \textbf{W} \Vert^2 = K$. 

The block diagram of a point-to-point communication system with a hybrid beamformer is shown in Fig. \ref{fig:Pointtopoint}. A hybrid beamformer consists of a baseband precoder $\textbf{F}_{\text{B}} \in \mathbb{C}^{M \times K}$ connected through $M$ RF chains to the RF beamformer $\textbf{F}_{\text{RF}} \in \mathbb{C}^{N_\text{t} \times M}$ such that $\textbf{F}=\textbf{F}_{\text{RF}}\textbf{F}_{\text{B}}$. The elements of the RF beamformer are either analog or digital $B$-bit resolution phase shifters as
\begin{align}
\textbf{F}_\text{RF}=  \begin{pmatrix}
  \text{e}^{j \theta_{11}} & \text{e}^{j \theta_{12}} & \cdots & \text{e}^{j \theta_{1M}} \\
  \text{e}^{j \theta_{21}} & \text{e}^{j \theta_{22}} & \cdots & \text{e}^{j \theta_{2M}} \\
  \vdots  & \vdots  & \ddots & \vdots  \\
\text{e}^{j \theta_{N_\text{t}1}} & \text{e}^{j \theta_{N_\text{t}2}} & \cdots & \text{e}^{j \theta_{N_\text{t}M}}
 \end{pmatrix},\: \: \: \forall \: \theta_{n_\text{t}m} \in {\Theta}, \: n_\text{t} \in \lbrace 1,\:...,\: N_\text{t}\rbrace, \: m \in \lbrace 1,\:...,\: M\rbrace, 
\end{align}
where $\Theta=[0,2\pi]$ for analog phase shifters and $\Theta=\lbrace 0,\: 2\pi /2^B, \: ...,\: (2^{B}-1)2\pi /2^B \rbrace$ for digital phase shifters. For the sake of the notation simplicity, throughout the paper it is assumed that $n_\text{t} \in \lbrace 1,\:...,\: N_\text{t}\rbrace$, $m \in \lbrace 1,\:...,\: M\rbrace$ and $k \in \lbrace 1,\:...,\: K\rbrace$ and the number of the RF chains at the transmitter and receiver are equal to $M$. Similar notation is used for the hybrid beamformer at the receiver as $\textbf{W}=\textbf{W}_{\text{RF}}\textbf{W}_{\text{B}}$ where $\textbf{W}_{\text{RF}} \in \mathbb{C}^{N_\text{r} \times M}$ and $\textbf{W}_{\text{B}} \in \mathbb{C}^{M \times K}$. Finally, the system model (\ref{eq:GeneralIOPointtoPoint}) for the hybrid scenario becomes
\begin{equation}
\hat{\textbf{s}}=\sqrt{\dfrac{P_\text{t}}{\Gamma_\text{t}\Gamma_\text{r}}}\textbf{W}_{\text{B}}^\text{H} \textbf{W}_{\text{RF}}^\text{H}\textbf{H}\textbf{F}_{\text{RF}}\textbf{F}_{\text{B}}\textbf{P}^{1/2}\textbf{s}+\sqrt{\frac{1}{\Gamma_\text{r}}}\textbf{W}_{\text{B}}^\text{H} \textbf{W}_{\text{RF}}^\text{H}\textbf{z} .
\end{equation}


In this paper, we derive an asymptotically optimal hybrid beamformer for a narrowband system under rich and sparse scattering channels under the assumption that E$[\Vert \textbf{H} \Vert^2]=N_\text{r}N_\text{t}$. Rayleigh fading with i.i.d. elements ${H}_{n_\text{r}n_\text{t}} \sim \mathcal{CN}(0,1)$ is employed to model the rich scattering channel. 

A geometry based model with $L \ll \min (N_\text{t},N_\text{r})$ multipath components is applied for the sparse scattering scenario. In this case, the channel matrix is expressed as \cite{Tse2005Book}
\begin{equation}
\label{eq:P2PChannel}
\textbf{H}=\sqrt{\dfrac{N_\text{t} N_\text{r}}{L}}\sum_{l=1}^L  \beta_l\textbf{a}_\text{r}(\phi_{\text{r}l})\textbf{a}_\text{t}^\text{H}(\phi_{\text{t}l}),
\end{equation}
where $\beta_l\sim \mathcal{CN}(0,1)$ is the multipath coefficient, $\phi_{\text{t}l}$ and $\phi_{\text{r}l}$ are angle-of-departure and angle-of-arrival of the $l$th multipath. Without loss of generality, it is assumed that $\vert \beta_1 \vert \geq  \vert \beta_2 \vert  \geq ... \geq \vert \beta_L \vert $. The steering vector $\textbf{a}_u(\phi_{ul})$, $\forall u \in \lbrace \text{t, r} \rbrace$, for linear arrays is expressed as
\begin{align}
\label{Eq:Manifoldvector}
\textbf{a}_u(\phi_{ul})=\dfrac{1}{\sqrt{N_u}}(1, \text{e}^{\frac{j2\pi d_u}{\lambda} \cos(\phi_{ul})}\: ...,\: \text{e}^{\frac{j2\pi d_u}{\lambda}(N_u-1)\cos(\phi_{ul})})^\text{T}
\end{align}
where $\phi_{ul}\in [0,\: \pi]$, $\lambda$ is the wavelength and $d_u$ is the antenna spacing \cite{Tse2005Book}. In the rest of this paper, it is assumed that the transmitter and the receiver are equipped with linear arrays with $d_u=\lambda/2$. 

\section{Problem Statement and Motivation}
\label{sec:Problem}
The optimal beamforming and power allocation for a fully digital point-to-point system is achieved by singular value decomposition (SVD) and waterfilling. The SVD factorizes the channel matrix as $\textbf{H}=\textbf{U} \boldsymbol{\Sigma}\textbf{V}^\text{H}$ where the columns of $\textbf{V}\in \mathbb{C}^{N_\text{t} \times N_\text{t} }$ and $\textbf{U}\in \mathbb{C}^{N_\text{r} \times N_\text{r} }$ are the right and left singular vectors of \textbf{H}, and the diagonal elements of $\boldsymbol{\Sigma}\in \mathbb{R}^{N_\text{r}  \times N_\text{t}}$ are the singular values of \textbf{H}. For a full-ranked \textbf{H}, the capacity of the MIMO channel at high SNR grows linearly with min$(N_\text{t},\: N_\text{r}) $ when $K=\min (N_\text{t},\: N_\text{r})$ streams are transmitted over the channel \cite{Goldsmith2003}. When $K \leq \min (N_\text{t},\: N_\text{r})$, the maximum achievable rates are derived by setting the combining and precoding matrices based on thin-SVD as $\textbf{W}_\text{d}=\textbf{U}_{1:K}$ and $\textbf{F}_\text{d}=\textbf{V}_{1:K}$ \cite{MOlischAntennaselection2005Journal}. In this case, $\Gamma_\text{t}=\Gamma_\text{r}=1$ and the capacity of a point-to-point system with $K$ streams over $\textbf{H}$ with Gaussian entries $s_k$ is \cite{MOlischAntennaselection2005Journal}
\begin{align}
\label{eq:Hcapacity}
C&=\max I(\textbf{s};\hat{\textbf{s}})=\max_{\text{trace}(\textbf{P})\leq 1} \log_2  \text{det} \Big ( \textbf{I}_K+\dfrac{P_\text{t}}{\sigma_z^2}\textbf{R}_\text{n}^{-1} \textbf{W}_\text{d}^\text{H}\textbf{H} \textbf{F}_\text{d} \textbf{P}\textbf{F}_\text{d}^\text{H} \textbf{H}^\text{H} \textbf{W}_\text{d} \Big ) \\ \nonumber
&=\max_{\sum_{k=1}^{K}P_{kk} \leq1 }  \sum_{k=1}^K \log_2 (1+P_\text{t} P_{kk} \sigma_{kk}^2 /\sigma_z^2)
\end{align}
where $I(\textbf{s};\hat{\textbf{s}})$ is the mutual information between $\textbf{s}$ and $\hat{\textbf{s}}$, $\textbf{R}_\text{n}=\frac{1}{\Gamma_\text{r}}\textbf{W}^\text{H}\textbf{W}=\frac{1}{\Gamma_\text{r}}\textbf{W}_\text{d}^\text{H}\textbf{W}_\text{d}=\textbf{I}_K$, $\sigma_{kk}^2$ are the ordered eigenvalues of $\textbf{H} \textbf{H}^\text{H}$ and the optimal $P_{kk}$ is derived by waterfilling. In this case, the capacity growth at high SNR is proportional to $K$. It should be noted that if the channel is rank-deficient it is not possible to transmit more than $\text{rank}(\textbf{H})$ symbols. When the hybrid beamformers are used, the achievable rate is expressed as \cite{Goldsmith2003} 
\begin{align}
\label{eq:AchievableRate1}
R&=I(\textbf{s};\hat{\textbf{s}}) =\log_2  \text{det} \Big ( \textbf{I}_K+\frac{\rho}{\Gamma_\text{t}\Gamma_\text{r}} \textbf{R}_\text{n}^{-1}\textbf{W}_\text{B}^\text{H}\textbf{W}_\text{RF}^\text{H}\textbf{H} \textbf{F}_\text{RF} \textbf{F}_\text{B} \textbf{P}\textbf{F}_\text{B}^\text{H}\textbf{F}_\text{RF}^\text{H} \textbf{H}^\text{H}\textbf{W}_\text{RF}\textbf{W}_\text{B}\Big ), 
\end{align}
where $\rho=\frac{P_\text{t}}{\sigma_z^2}$ is a measure of link signal-to-noise ratio (SNR). 

One of the main challenges of designing hybrid beamformers is the joint design of the RF beamformers and baseband precoders/combiners considering the constant modulus constraint on the phase shifters. Designing $\textbf{F}_\text{B},\: \textbf{F}_\text{RF},\: \textbf{W}_\text{B}$ and $ \textbf{W}_\text{RF}$ to maximize (\ref{eq:AchievableRate1}) is a nonconvex problem and in general it is difficult to solve \cite{SpatiallySparsePrecodingAyach}. Due to the similarity between the hybrid beamformer matrices at the transmitter and the receiver, same design algorithms are applicable to both sides. Hence, the discussions and derivations during this paper are mostly focused on the hybrid beamformer at the transmitter. In this case, it is desired to maximize the mutual information $I(\textbf{s};\textbf{y})$ as 
\begin{align}
\label{eq:OptimalJointDesign}
(\textbf{F}_\text{B}^\text{opt} , \textbf{F}_\text{RF}^\text{opt})=\underset{\textbf{F}_\text{B}, \textbf{F}_\text{RF}}{\arg\max} \: I(\textbf{s};\textbf{y})=\underset{\textbf{F}_\text{B}, \textbf{F}_\text{RF}}{\arg\max} \: \log_2  \text{det} \Big ( \textbf{I}_{N_\text{r}}+\frac{\rho}{\Gamma_\text{t}} \textbf{H} \textbf{F}_\text{RF} \textbf{F}_\text{B} \textbf{P}\textbf{F}_\text{B}^\text{H}\textbf{F}_\text{RF}^\text{H} \textbf{H}^\text{H}\Big ),
\end{align}
subject to (s.t.) $\vert F_{\text{RF},n_\text{t}m} \vert =1$, where $\textbf{F}_\text{B}^\text{opt}$ and $\textbf{F}_\text{RF}^\text{opt}$ are the optimal baseband precoding and RF beamforming matrices. It has been shown that based on some approximations, the optimization in (\ref{eq:OptimalJointDesign}) can be reformulated as \cite{SpatiallySparsePrecodingAyach}
\begin{align}
\label{eq:FoptProblem}
(\textbf{F}_\text{B}^\text{opt} , \textbf{F}_\text{RF}^\text{opt})=\underset{\textbf{F}_\text{B}, \textbf{F}_\text{RF}}{\arg\min} \Vert \textbf{F}_\text{d} - \sqrt{\frac{1}{\Gamma_\text{t}}}\textbf{F}_\text{RF} \textbf{F}_\text{B} \Vert, \:\:\:\: \text{ s.t. }\vert F_{\text{RF},n_\text{t}m} \vert =1.
\end{align}
A suboptimal joint baseband and RF design based on matching pursuit was proposed to solve (\ref{eq:FoptProblem}) for a sparse scattering channel \cite{SpatiallySparsePrecodingAyach}. For a more general channel, including Rayleigh fading, this optimization problem can be approximated as a convex problem and a joint iterative suboptimal solution was proposed in \cite{NiDL15}. Another approach to design the hybrid beamformer is to calculate $\textbf{F}_\text{RF}^\text{opt}$ at the first step, and then derive the $\textbf{F}_\text{B}^\text{opt}$ for the effective channel $\textbf{H}_\text{e}  =\textbf{HF}_\text{RF}$. Letting $\textbf{x}_\text{B}=\textbf{F}_\text{B}\textbf{P}^{1/2}\textbf{s}$, data-processing inequality indicates that \cite{Cover:2006:EIT:1146355}
\begin{equation}
 I(\textbf{s};\textbf{y}) \stackrel{(a)}{\leq}  I(\textbf{x}_\text{B};\textbf{y}) \leq C.
\end{equation}
where inequality $(a)$ turns into equality when $\textbf{F}_\text{B}= \textbf{V}_\text{e}$ as $\textbf{H}_\text{e}=\textbf{HF}_\text{RF}=\textbf{U}_\text{e}\boldsymbol{\Sigma}_\text{e} \textbf{V}^\text{H}_\text{e}$, and \textbf{P} is derived by waterfilling. It could be concluded that max $I(\textbf{s};{\textbf{y}})$ only depends on the design of $ \textbf{F}_\text{RF}$. In this case,
\begin{align}
\label{eq:AchievableRate4}
\textbf{F}_\text{RF}^\text{opt} =\underset{ \textbf{F}_\text{RF}}{\arg\max} \: I(\textbf{s};\textbf{y})=\underset{ \textbf{F}_\text{RF}}{\arg\max} \: \log_2  \text{det} \Big ( \textbf{I}_{N_\text{r}}+\frac{\rho}{\Gamma_\text{t}} \textbf{H} \textbf{F}_\text{RF}  \textbf{V}_\text{e} \textbf{P} \textbf{V}_\text{e}^\text{H}\textbf{F}_\text{RF}^\text{H} \textbf{H}^\text{H}\Big ) , \:\:\:\:\text{ s.t. }\vert F_{\text{RF},n_\text{t}m} \vert =1,
\end{align}
where $\Gamma_\text{t}=\text{trace}(\textbf{F}_\text{RF}\textbf{V}_\text{e}\textbf{V}^\text{H}_\text{e} \textbf{F}_\text{RF}^\text{H})/K=\text{trace}(\textbf{F}_\text{RF} \textbf{F}_\text{RF}^\text{H})/K=N_\text{t}$. The two-stage design of $\textbf{F}_\text{B}$ and $\textbf{F}_\text{RF}$ has been previously studied in \cite{PerAntennaPi2012}-\cite{FoadTorontoApril2015Digital}. However, the spectral efficiency based on these works depends on numerical calculations and it is not possible to derive the closed-form expression of the performance. Based on the two-stage approach, a virtually optimal hybrid beamforming and the closed-form expression of the spectral efficiency for a point-to-point system with large number of antennas under two specific channel scenarios are presented in the next section.

\section{Hybrid Beamforming for the Point-to-Point Scenario}
\label{sec:proposedalgp2p}

In this section, an asymptotically optimal hybrid beamformer that maximizes the achievable rate in (\ref{eq:OptimalJointDesign}) is presented. Initially, based on some basic properties of the elements of the singular vectors, it will be shown that analog phase shifters with $K=M/2$ can achieve the performance of digital beamformers. It is notable that the analysis presented for this scenario is a modification of the approach in \cite{MOlischAntennaselection2005Journal}. Under this assumption the system is underperforming as the multiplexing gain is limited to $M/2$. In order to develop a hybrid beamforming algorithm that efficiently employs all the RF chains to transmit $K=M$ streams, some of the properties the singular vectors of \textbf{H} are investigated. Then, the hybrid beamforming solution for large antenna array systems with analog phase shifters are presented. For the case of $K<M<2K$, a combination of the methods for $M=K$ and $M=2K$, and its performance is discussed. When digital phase shifters are employed, a simple heuristic suboptimal solution and its performance lower-bound is presented. Finally, a discussion on the proposed method and a comparison with the state-of-the-art are provided at the end of the section.

Since \textbf{V} is a unitary matrix, $\textbf{v}_k^\text{H} \textbf{v}_k=\sum_{n_\text{t}=1}^{N_\text{t}} \vert V_{n_\text{t}k} \vert^2=1$ and $\vert V_{n_\text{t}k} \vert \leq 1$. Thus $\vert V_{n_\text{t}k} \vert$ is in the domain of the inverse cosine function. Hence,
\begin{align}
\vert V_{n_\text{t}k }\vert \text{e}^{j\angle V_{n_\text{t}k }} &=\text{e}^{j\angle V_{n_\text{t}k }}\cos \Big(\cos^{-1} (\vert V_{n_\text{t}k } \vert)\Big) = \frac{\text{e}^{j\angle V_{n_\text{t}k }}}{2}\text{e}^{j\cos^{-1} (\vert V_{n_\text{t}k } \vert)}+\frac{\text{e}^{j\angle V_{n_\text{t}k }}}{2}\text{e}^{-j\cos^{-1} (\vert V_{n_\text{t}k } \vert)} \\ \nonumber
&=\frac{1}{2}\text{e}^{j\angle V_{n_\text{t}k } + j\cos^{-1} (\vert V_{n_\text{t}k } \vert)}+\frac{1}{2}\text{e}^{j\angle V_{n_\text{t}k } - j\cos^{-1} (\vert V_{n_\text{t}k } \vert) }.
\end{align}
This means that two phase shifters and an adder at the RF are sufficient to produce $V_{n_\text{t}k}$ when $M=K$ RF chains and $2MN_\text{t}$ phase shifters are available. Alternative approach to the adders is employing $M=2K$ RF chains and $2MN_\text{t}$ phase shifters. In this case, $1/\sqrt{\Gamma_\text{t}}\textbf{F}_\text{RF}^\text{opt} \textbf{F}_\text{B}^\text{opt}=\textbf{F}_\text{d} $ is achieved by setting
\begin{align}
\label{eq:VphaseFind}
 F_{\text{RF},n_\text{t}k'}^\text{opt}=
\begin{cases}
 &  \text{e}^{j\angle V_{n_\text{t}k } + j\cos^{-1} (\vert V_{n_\text{t}k } \vert)} \text{ for } k'=2k-1 \\ 
 &  \text{e}^{j\angle V_{n_\text{t}k } - j\cos^{-1} (\vert V_{n_\text{t}k } \vert) } \text{ for } k'=2k,\\
  \end{cases}
\end{align}
and $\textbf{F}_\text{B}^\text{opt}=\frac{1}{2}\text{diag}(\textbf{1}_{2\times 1}, \: ..., \: \textbf{1}_{2\times 1})$ and $\Gamma_\text{t}=1$. Hence, the maximum rate in (\ref{eq:Hcapacity}) can be achieved with this design. In order to derive $\textbf{F}_\text{B}^\text{opt}$ and $\textbf{F}_\text{RF}^\text{opt}$ for $M=K$ scenario, further properties of the singular vectors are investigated in the following subsection.

\subsection{Properties of the Channel Singular Vectors}
The behaviors of the channel singular vectors for Rayleigh and geometry based models are presented in Theorem 1 and Lemma 1 in the following.

\textit{Theorem 1}: If $H_{n_\text{r}n_\text{t}} \sim \mathcal{CN}(0,1)$ and $N_\text{t}\to \infty$, $N_\text{r}\to \infty$, then the elements of the singular vectors of \textbf{H} are i.i.d and follow $\sqrt{N_\text{t}} V_{n_\text{t}n_\text{t}'}, \: \sqrt{N_\text{r}}{U}_{n_\text{r}n_\text{r}'}  \sim \mathcal{CN}(0,1)$, $\forall \: n_\text{t}, n_\text{t}' \in \lbrace 1, \: ...,\: N_\text{t} \rbrace$ and $\forall \: n_\text{r},  n_\text{r}' \in \lbrace 1, \: ...,\: N_\text{r} \rbrace$.

\textit{Proof}: It is known that the left and right singular vectors of $\textbf{H}=\textbf{U} \boldsymbol{\Sigma}\textbf{V}^\text{H}$ are uniformly distributed on a complex $N_\text{t}$-hypersphere and a $N_\text{r}$-hypersphere with radius 1 \cite{GrassmanianBF}. As a result, $\sqrt{N_\text{t}}\textbf{v}_{n_\text{t}}$ and $\sqrt{N_\text{r}}\textbf{u}_{n_\text{r}}$ are uniformly distributed on the surface of a $N_\text{t}$ and $N_\text{r}$ dimensional hyperspheres with radius $\sqrt{N_\text{t}}$ and $\sqrt{N_\text{r}}$. Moreover, the coordinates of a randomly chosen point according to a uniform distribution on an $N$-hypersphere of radius $\sqrt{N}$ are i.i.d. with $\mathcal{CN}(0,1)$ when $N \to \infty$ \cite{ECP1294}. Hence, the elements of $\sqrt{N_\text{t}}\textbf{v}_{n_\text{t}}$ and $\sqrt{N_\text{r}}\textbf{u}_{n_\text{r}}$ are i.i.d. with $\mathcal{CN}(0,1)$. \hfill\(\Box\)

\textit{Remark:} As far as the authors are aware, the distribution of the elements of the singular vectors of matrix \textbf{H}, when  $H_{n_\text{r}n_\text{t}}\sim \mathcal{CN}(0,1)$ for $N_\text{t} \to \infty$ and $N_\text{r} \to \infty$, has not been previously reported in the literature, although the pieces of the proof have been available for a long time and they have been studied by different researcher such as Love and Spruill \cite{GrassmanianBF,ECP1294}. 

\begin{figure}
\centering
\includegraphics[width=4 in]{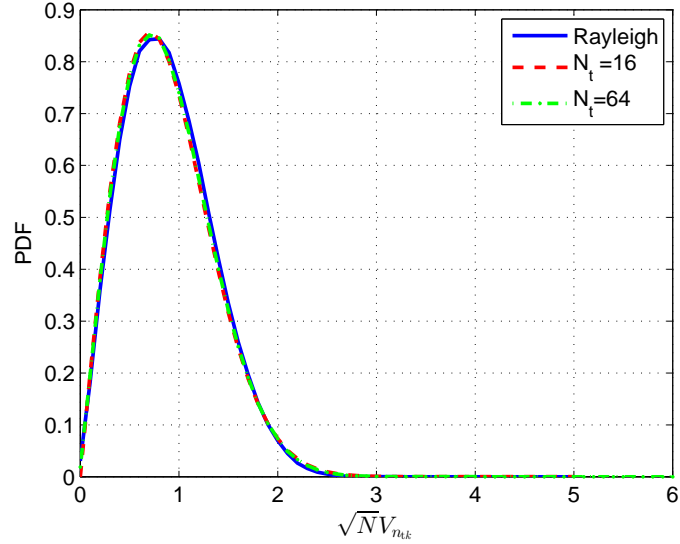}
\caption{Comparison between the probability density function (PDF) of $\sqrt{N_\text{t}}\vert V_{n_\text{t}k} \vert$ when it follows Rayleigh distribution with parameter $\sigma_\text{R}= \frac{1}{\sqrt{2}}$, simulation results for the PDF of $\sqrt{N_\text{t}}\vert V_{n_\text{t}k} \vert$ for a Rayleigh fading channel over 1000 realizations with $N_\text{t}=N_\text{r}=16$ and $N_\text{t}=N_\text{r}=64$.}
\label{fig:SingularVecAmpRayleigh}
\end{figure}

The real and imaginary parts of random variables with $\mathcal{CN}(0,1)$ are distributed as $\mathcal{RN}(0,\frac{1}{2})$ \cite{Tse2005Book}. Hence, $\sqrt{N_\text{t}} \vert V_{n_\text{t}k} \vert$ has a Rayleigh distribution with parameter $\sigma_\text{R}= \frac{1}{\sqrt{2}}$ and its expected value is $\sigma_\text{R}\sqrt{\frac{\pi}{2}}$ \cite{citeulike:9070763}. Fig. \ref{fig:SingularVecAmpRayleigh} shows that the Rayleigh distribution can provide a good approximation even for a finite $N_\text{t}\in \lbrace 16, \: 64 \rbrace$. The properties of the sparse scattering channels are described in the following lemma.

\textit{Lemma 1} \cite{BeamsteeringAyach2012}: For a geometry based channel model with $N_\text{t}\to \infty$ and $N_\text{r}\to \infty$, the relationship between the singular and steering vectors is expressed as $\textbf{v}_l=\textbf{a}_\text{t}(\phi_{\text{t}l})$ and $\textbf{u}_l=\textbf{a}_\text{r}(\phi_{\text{r}l})$, $\forall \: l \in \lbrace 1, \: ...,\: L \rbrace$. 

\subsection{Hybrid Beamforming for $M=K$ Scenario}
\label{sec:P2PAlgs}
The proposed hybrid beamformer when $M=K$ is presented in the following lemma.

\textit{Lemma 2:} The asymptotically optimal solution $\textbf{F}_\text{RF}^\text{opt}$ and $\textbf{W}_\text{RF}^\text{opt}$ to the optimizations in (\ref{eq:AchievableRate1}), (\ref{eq:OptimalJointDesign}) and (\ref{eq:AchievableRate4}) for large $N_\text{t}$ and $N_\text{r}$ with $M=K$ and analog phase shifters is $F_{\text{RF},n_\text{t}k}^\text{opt}=\text{e}^{j\angle V_{n_\text{t}k}}$, $W_{\text{RF},n_\text{r}k}^\text{opt}=\text{e}^{j\angle U_{n_\text{r}k}}$. In this case, the baseband precoder and combiner matrices become $\textbf{F}_\text{B}=\textbf{W}_\text{B}=\textbf{I}_K.$

\textit{Proof:} Refer to Appendix \ref{App:lemma2}. \hfill\(\Box\)

It was previously shown that for the geometry based channel models, (\ref{eq:OptimalJointDesign}) could be approximated by (\ref{eq:FoptProblem}) which is equivalent to minimizing the Euclidean distance between $\textbf{F}_\text{d}$ and $1/\sqrt{N_\text{t}}\textbf{F}_{\text{RF}}\textbf{F}_{\text{B}}$ \cite{SpatiallySparsePrecodingAyach}. In Appendix \ref{App:AlternativeProof}, it is proved that the proposed RF beamformer of Lemma 2 can be alternatively derived by 
\begin{align}
\label{eq:RFOpt1}
\underset{\textbf{F}_{\text{RF}}}{\text{minimize}}  \quad  \Vert \sqrt{\frac{1}{N_\text{t}}} \textbf{F}_{\text{RF}}- \textbf{F}_\text{d} \Vert^2,  \:\:\:\: \text{s.t. }\vert F_{\text{RF},n_\text{t}k}\vert=1.
\end{align}

In order to implement the hybrid beamformer of Lemma 2, the first $K$ singular vectors and values of \textbf{H} should be initially calculated. Then, each phase shifter at the transmitter and the receiver is directly set to the phase of the corresponding element in the right and left singular vectors, respectively. Considering the impact of the RF beamformers, the baseband precoder and combiner matrices are equal to an identity matrix. Finally, the optimal allocated power to each symbol is derived by waterfillining. The performance of the proposed hybrid beamformer compared to $C$ in (\ref{eq:Hcapacity}) for $M=K$ and Rayleigh fading channel is expressed in the following lemma.

\textit{Lemma 3}: For large $N_\text{t}$ and $N_\text{r}$ and at high SNR regime, the difference between the maximum rate $C$ form (\ref{eq:Hcapacity}) and the rate $R_\text{C}$ achieved by the beamforming scheme of Lemma 2 for a Rayleigh channel is expressed as
\begin{equation}
\label{eq:lemma3}
C-R_C=-2K\log_2 (\frac{\pi}{4}).
\end{equation}

\textit{Proof}: Refer to Appendix \ref{App:Lemma3Proof}.  \hfill\(\Box\)

Lemma 3 indicates that the spectral efficiency achieved by the digital beamformers is 0.7$K$ bits/Hz/s more than the performance of the hybrid beamformers when the number of antennas are large and the channel is modeled by Rayleigh fading.

When $K<M < 2K$, two RF chains are used per symbol to transmit in the direction of the singular vectors corresponding to the first $M-K$ singular values of the channel. In this case, the hybrid beamformer of $M=2K$ is used. For the remaining $2K-M$ symbols, the hybrid beamformer of Lemma 2 is used. That is $2(M-K)$ RF chains are used to transmit $M-K$ symbols and each of the remaining $2K-M$ symbols are transmitted on one of the remaining RF chains. For example, assuming that $K=3$ and $M=5$, the baseband precoder becomes $\textbf{F}_\text{B}^\text{opt}=\text{diag}(\frac{1}{2}\textbf{1}_{2\times 1}, \: \frac{1}{2}\textbf{1}_{2\times 1}, \: 1)$. Then, (\ref{eq:VphaseFind}) is used to design the the RF beamforming vectors $\textbf{f}_{\text{RF},1}, \: \textbf{f}_{\text{RF},2}$ according to $\textbf{v}_1$, and $\textbf{f}_{\text{RF},3}, \: \textbf{f}_{\text{RF},4}$ based on $\textbf{v}_2$. Finally, $\textbf{f}_{\text{RF},5}$ is adjusted based on $\textbf{v}_3$ and Lemma 2. Similar approach can be also applied at the receiver side. In a general scenario that $K \leq M \leq 2K$, by following the results of Appendix \ref{App:lemma2} and \ref{App:Lemma3Proof}, it can be easily verified that (\ref{eq:lemma3}) becomes
\begin{equation}
\label{eq:generalMandK}
C-R_C=-2(2K-M)\log_2 (\frac{\pi}{4}).
\end{equation}
For example, letting $K=3$ and $M=5$, then $C-R_C=-2\log_2 (\frac{\pi}{4})$. It should be noted that adding an extra RF chain at each side can increase the spectral efficiency by $-\log_2 (\frac{\pi}{4})$. However, this improvement will also increase the system cost, complexity and power consumption.

For a geometry based channel, the singular vectors and the steering vectors become equal and the proposed algorithm will be translated into steering the beams towards the channel multipath components as proposed in \cite{BeamsteeringAyach2012}. Following a similar approach as in Appendix \ref{App:Lemma3Proof}, it can be easily shown that $1/\sqrt{N_\text{t}}\textbf{a}_\text{t}(\phi_{\text{t},k})^\text{H}\textbf{f}_\text{RF}=1$ and $C-R_\text{C}=0$. Hence, extra RF chains $M-K>0$ will not improve the performance in such channels.

\subsection{Digital Phase Shifters}
Another challenge for designing hybrid beamformers is the discrete resolution of the phase shifters. When $B$-bit resolution phase shifters are employed, the search space for the optimum set of phases becomes $2^{BMN_\text{t}}$ which can be very large for large $N_\text{t}$. As an example, when there are $N_\text{t}=64$, $M=4$ and 2-bit resolution phase shifters, there are $2^{512}$ possible phase combination which is computationally expensive to search in the real-time applications. One way out is the use a predefined set of phases known as RF codebooks \cite{SpatiallySparsePrecodingAyach,SohailICC2015}. The disadvantage of the RF codebooks is that they are usually designed for a fixed type of channel such as sparse channels. The alternative approach to design the RF beamformer with discrete resolution phase shifters is rounding the phases as
\begin{align}
\label{eq:SolutionDiscreteres}
\theta_{n_\text{t}k}^\text{d}= \arg \min_{\theta_{n_\text{t}k}} \:  \vert \angle F_{\text{d},n_\text{t}k} -\theta_{n_\text{t}k} \vert ,\:\:\:\: \text{s.t. } \theta_{n_\text{t}k}\in \lbrace 0,\: ...,\: (2^{B}-1)2\pi /2^B \rbrace ,
\end{align}
where $\theta_{n_\text{t},k}^\text{d}$ is the phase of $F_{\text{RF},n,k}$. The lower-bound on the rate loss with this design is provided in the following lemma.

\textit{Lemma 4}: The gap between $R_\text{C}$ and the achievable rate $R_\text{D}$ by the hybrid beamformer based on (\ref{eq:SolutionDiscreteres}) with $B$-bit resolution digital phase shifters is bounded as
\begin{equation}
\label{eq:DisResGap}
R_\text{C} - R_\text{D} \leq  -K\log_2\Big(\cos^4 (\frac{2 \pi}{2^{B+1}}) \Big).
\end{equation}
\textit{Proof}: Refer to Appendix \ref{App:DisAlg2}. \hfill\(\Box\)

Lemma 4 indicates that hybrid beamformers with analog phase shifters can achieve maximum $0.45K$ bits/Hz/s higher spectral efficiency compared to the scenario that digital phase shifters with $B=3$ are employed. As hybrid beamformers target the transmission of a small number of symbols, the gains achieved by using analog phase shifters are negligible at high SNR regime. In addition, the low cost and computational complexity of the proposed scheme in (\ref{eq:SolutionDiscreteres}) makes it an effective approach for practical applications.

\subsection{Discussion and Comparison with the State-of-the-Art}

In this paper, the analytical discussions are focused on asymptotically large antenna arrays. This is in contrast to the works in \cite{SpatiallySparsePrecodingAyach,SohailICC2015,NiDL15,PerAntennaPi2012,FoadTorontoApril2015Digital,NiDL15} where the analysis are presented for limited number of antennas. The advantages of considering asymptotically large arrays are two-fold. Firstly, it facilitates the analysis to derive the virtually optimal hybrid beamformer and the closed-forms for the achievable spectral efficiency. Secondly, as it will be shown in section VII, the simulation results indicate that the analysis for the asymptoticly large array scenario provides a reliable estimate of achievable performance for scenarios with limited number of antennas.

One of the common approaches in the literature is to decompose the unconstrained thin-SVD based beamformer matrix into RF beamformer and baseband precoder matrices, \cite{SpatiallySparsePrecodingAyach,SohailICC2015,NiDL15}. The computational complexity of the rank-$M$ thin-SVD of \textbf{H} is $O(N_\text{t} N_\text{r} M)$ for $M \ll \sqrt{N_\text{t} N_\text{r}}$ \cite{Brand200620}. The state-of-the-art hybrid beamformers that require a second round of computations to decompose $\textbf{F}_\text{d}$ into $\textbf{F}_\text{B}^\text{opt}$ and $\textbf{F}_\text{RF}^\text{opt}$ can cause high computational delay and complexity \cite{SpatiallySparsePrecodingAyach,SohailICC2015}. An iterative algorithm can be used to solve the optimization problem in (\ref{eq:OptimalJointDesign}), however, the iterative algorithm renders a high computational cost and delay \cite{PerAntennaPi2012,FoadTorontoApril2015Digital,NiDL15}. For example, the complexity of the hybrid beamformer in \cite{FoadTorontoApril2015} is $O(\max (N_\text{t},N_\text{r})^2 \min(N_\text{t},N_\text{r}))$. 

Compared to the state-of-the-art, the proposed hybrid beamformer of lemma 2 is faster and it is virtually the optimal scheme for the systems with large $N_\text{t}$, $N_\text{r}$ operating in Rayleigh and sparse channels. The computational complexity of the proposed scheme is equal to the complexity of rank-$M$ thin-SVD as $O(N_\text{t} N_\text{r} M)$. In addition, the closed-form expressions of the achievable rates are derived which to best of the authors' knowledge was not previously reported.

\section{Multiuser Scenario}
\label{sec:MU}
In the downlink scenario, the base station with $N_\text{t}$ antennas transmits $K$ symbols $\textbf{s} \in \mathbb{C}^{K \times 1}$ to $K$ single antenna mobile stations where E$[\textbf{ss}^\text{H}]=1/K\textbf{I}_{K}$. In this scenario, it is assumed that the base station has perfect CSI and the users cannot collaborate. The total transmit power and the wireless channel matrix are denoted as $P_\text{t}$ and $\textbf{H}\in \mathbb{C}^{K \times N_\text{t}}$, respectively. The transmit vector is expressed as $\textbf{x}= \sqrt{P_\text{t}/\Gamma_\text{t}} \textbf{Fs}$ where \textbf{F} is the precoding matrix and 
\begin{equation}
\label{eq:MUGamma}
\Gamma_\text{t}=\text{E\Big[trace}(\textbf{F}\textbf{s}\textbf{s}^\text{H}\textbf{F}^\text{H})\Big]=\text{E}\Big[\text{trace}  (\textbf{F}\textbf{F}^\text{H}  )\Big] /K 
\end{equation}
is a power normalization factor. The channel output vector is $\textbf{y}=(y_1, \: ... ,\: y_K)^\text{T}$ where $y_k$ is the received signal at $k$th mobile station. The system input-output relation is expressed as $\textbf{y}=\sqrt{\frac{P_\text{t}}{\Gamma_\text{t}}} \textbf{HF} \textbf{s}+\textbf{z},$ where $\textbf{z}=(z_1, \: ... ,\: z_K)^\text{T}$, $\text{E}[\textbf{zz}^\text{H}]=\sigma_z^2\textbf{I}_K$ contains the receiver noise. The optimal sum-rate capacity of \textbf{H} is derived by \cite{WishvanathTse}
\begin{align}
\label{eq:MUCap}
C_\text{sum}(P_\text{t},\textbf{H})= \max_{\text{trace(\textbf{P})}\leq 1}\log_2 \text{det} \Big (\textbf{I}_K+\frac{P_\text{t}}{\sigma_z^2}\textbf{P} \textbf{H}\textbf{H}^\text{H} \Big),
\end{align}
where \textbf{P} is the power allocation matrix. In general, the capacity of the broadcast channels is derived by dirty paper coding which is difficult to implement \cite{Goldsmith2003}. Hence, in practice the suboptimal linear precoding algorithms with low complexity such as zero-forcing (ZF) are preferred. It has been shown that the performance of ZF converges to optimal sum-capacity for the Rayleigh channel when $N_\text{t}$ goes large \cite{MasiveMIMORusek}. For the hybrid structure, the vector of the received signals becomes $\textbf{y}=\sqrt{\frac{P_\text{t}}{\Gamma_\text{t}}} \textbf{HF}_\text{RF} \textbf{F}_\text{B} \textbf{s}+\textbf{z}.$ In the following lemma, we present the virtually optimal hybrid beamformer and its performance, achievable sum-rate with respect to the sum-rate capacity of \textbf{H}, for multiuser MIMO scenario when the channel is modeled by Rayleigh fading.

\textit{Lemma 5:} The asymptotically optimal hybrid beamformer for the multiuser scenario with Rayleigh channel and in the high SNR regime consists of $\textbf{F}_\text{RF}^\text{opt}$ from Lemma 2, and $\textbf{F}_\text{B}^\text{opt}=(\textbf{H}\textbf{F}_\text{RF}^\text{opt})^{-1}$. In this case, the difference between the sum-capacity $C_\text{sum}$ and the maximum achievable sum-rate $R_{\text{sum}}$ at high SNR is 
\begin{equation}
\label{eq:lemma5}
C_\text{sum}(P_\text{t},\textbf{H})-R_\text{sum}(P_\text{t},\frac{1}{\sqrt{\Gamma_\text{t}}}\textbf{H}\textbf{F}_\text{RF}^{\text{opt}}\textbf{F}_\text{B}^\text{opt})=-K\log_2 (\pi/4).
\end{equation}

\textit{Proof}: Refer to Appendix \ref{AppendixMU}. \hfill\(\Box\)

The difference between right hand side of (\ref{eq:lemma3}) and (\ref{eq:lemma5}) is a scalar number 2. This factor comes from the fact that the transmitter and the receiver in the point-to-point system are equipped with hybrid beamformer, and the losses imposed by the RF beamformer should be counted at both sides. 

For the case of the geometry based channels, the proposed RF beamformer in Lemma 2 is still asymptotically the optimal beamformer as it is shown in Appendix \ref{AppendixMU}. In order to achieve the maximum achievable rate, nonlinear precoding schemes should be used at the baseband precoder. The performance of ZF baseband precoder for sparse channels and the multiantenna multiuser scenario considering the impact of imperfect CSI on the system performance is investigated in \cite{HeathMUMIMO}. Under the assumption of single antenna users, sparse channel, a base station with a linear array and ZF baseband precoder, the beamformer of Lemma 5 and the algorithm in \cite{HeathMUMIMO} will result in the same performance. However, the hybrid beamformer in \cite{HeathMUMIMO} is not applicable to Rayleigh channels due to employing a special RF codebook. The RF beamformer of Lemma 5, however, is applicable to both rich and sparse channels, and it is adaptable to different scenarios. 

\section{Hybrid Beamforming with Phase Shifter Selection for Rayleigh Channel}
\label{sec:PSselect}
In the previous sections, it was shown that the spectral efficiency achieved by the hybrid beamformers with $MN_\text{t}$ phase shifters is comparable to the performance of the digital beamformers. However, the power consumption in the phase shifter network can be significant when large number of antennas are employed. Additionally, it is expected that the elements of $\textbf{F}_\text{RF}$ corresponding to the elements of $\textbf{F}_\text{d}$ with smaller amplitudes have less impact on the performance of the beamformer. In this case, there are phase shifters with insignificant contribution to spectral efficiency although they consume the same amount of power. For this reason, a novel phase shifter selection scheme that turns off those shifters according to a predefined threshold is proposed in this section. The advantages of such an approach are twofold. Firstly it can improve the spectral efficiency as more power will be transmitted through the phase shifters with more contribution. Secondly, it can reduce the power consumption of the phase shifter network. In general, switches consume less power compared to phase shifters. The power consumption of each of the phase shifters $P_\text{PS}$ and switches $P_\text{S}$ at 2.4 GHz are reported as $28.8\leq P_\text{S}\leq 152$ mW \cite{PhaseShifterSurvey} and $0<P_\text{S}\leq 15$ mW \cite{SwitchesSurvey}. By switching off $\beta\%$ of the phase shifters, the total consumed power in the RF beamformer becomes $P_\text{PSN}=MN_\text{t}((1-\beta/100)P_\text{PS}+P_\text{S})$. For example, if $P_\text{PS}=111$ mW, $P_\text{S}=1$ mW, $M=4$ and $N_\text{t}=64$, the consumed power in the RF beamformer with all the phase shifters in operation (without switches) and the scenario that $\beta =50$ are $MN_\text{t}P_\text{PS}=28.4$ W and $P_\text{PSN}=14.4$ W, respectively. That is close to $50\%$ power saving in the phase shifter network. In this case, $\textbf{F}_\text{RF}$ can be set as
\begin{align}
\label{eq:PSselectionBF}
{F}_{\text{RF},n_\text{t}k}= 
\begin{cases}
   0, &  \sqrt{{N}_\text{t}} \vert V_{n_\text{t}k} \vert \leq \alpha ,\\
   \text{e}^{j \angle V_{n_\text{t}k}}, &  \ \alpha < \sqrt{{N}_\text{t}} \vert V_{n_\text{t}k} \vert,
  \end{cases}
  \end{align}
where $\alpha$ is the threshold level. In Appendix \ref{Appendix:ProofLemma6}, it is shown that the relationship between $\alpha$ and $\beta$ of is expressed as $\alpha=\sqrt{-\text{ln}(1-\beta/100)}$. In addition, when the RF beamformer is set according to (\ref{eq:PSselectionBF}) the baseband precoder and combiner matrices become $\textbf{F}_\text{B}=\textbf{W}_\text{B}=\textbf{I}_K$. In the following lemma, the closed-form expression for the performance of the proposed phase shifter selection algorithm is presented.

\textit{Lemma 6}: In a Rayleigh channel and at high SNR and for large $N_\text{t}$, $N_\text{r}$, the spectral efficiency $R_\beta$ achieved by the proposed phase shifter selection scheme, when $\beta \%$ of the phase shifters are switched off, compared to $C$ is obtained from
\begin{equation}
\label{eq:RProPSSelectionLemma}
C-R_\beta =2K\log_2 (1-\beta/100)-4K \log \Big( \frac{\sqrt{\pi}}{2}+ \alpha \text{e}^{-\alpha^2} -   \frac{\sqrt{\pi}}{2}\text{erf}(\alpha) \Big),
\end{equation}
where $\alpha=\sqrt{-\text{ln}(1-\beta/100)}$.

\textit{Proof:} Refer to Appendix \ref{Appendix:ProofLemma6}.

The closed-form (\ref{eq:lemma3}) in Lemma 3 is a special case of (\ref{eq:RProPSSelectionLemma}) when $\beta=0$. As a consequence of switching off the phase shifters that have smaller impact on the spectral efficiency, more power can be allocated to the elements with higher impact. As a result, the choice of $\beta$ can lead to a higher spectral efficiency as the power consumption in the RF beamformer is reduced. The relationship between $\beta$ and $R_\beta $ is studied in more detail by computer simulations in the next section.

\section{Simulation Results}
\label{sec:imulations}
\begin{figure}
  \centering
  \begin{minipage}[b]{0.48\textwidth}
  \includegraphics[width=3.2in]{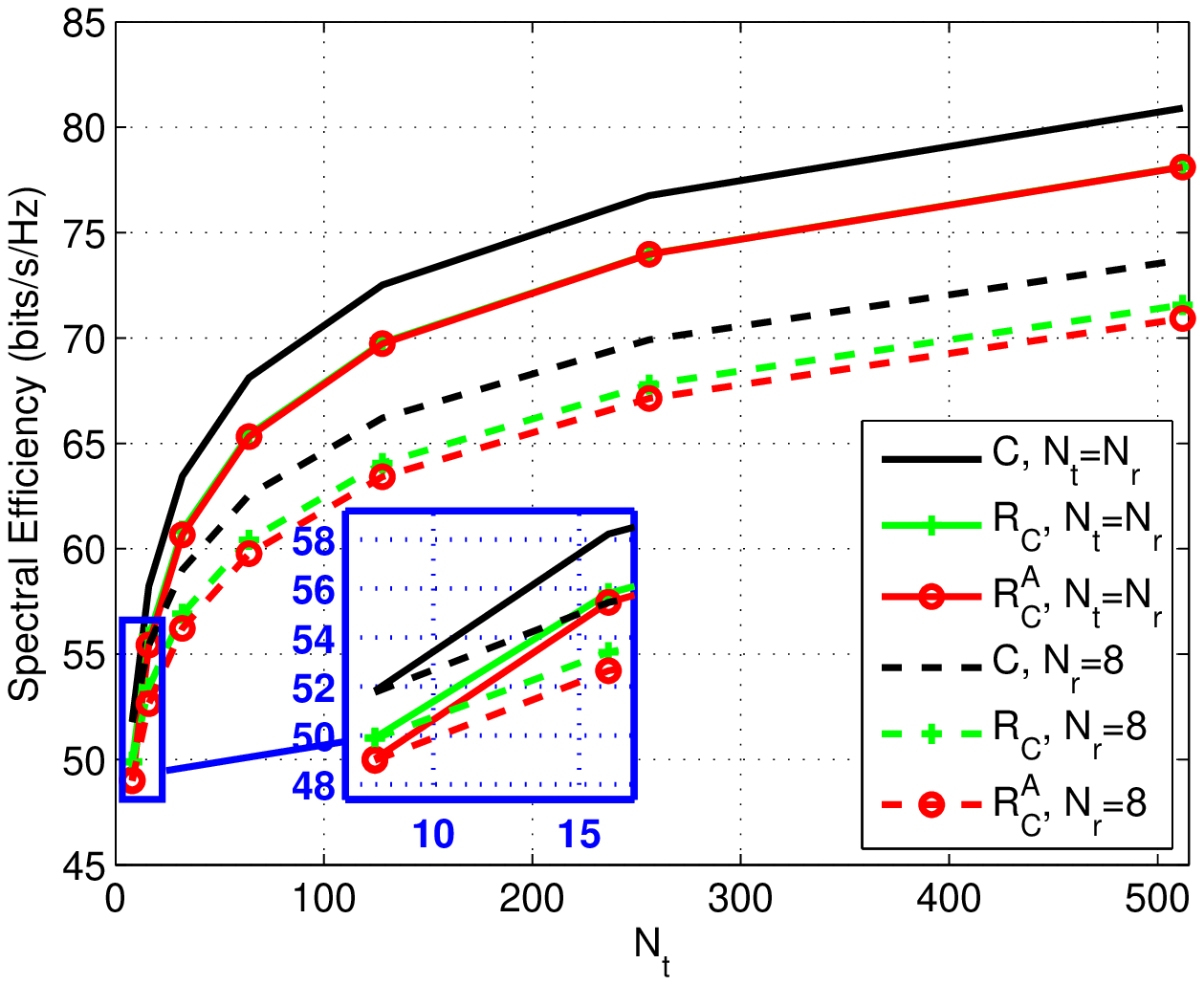}
\caption{$C, \: R_\text{C}, R_\text{C}^\text{A}$ when the number of the antennas varies, $\rho=34$ dB and Rayleigh channel.}
\label{fig:RayleighNumberofAntennas}
  \end{minipage}
  \hfill
  \begin{minipage}[b]{0.48\textwidth}
\includegraphics[width=3.2in]{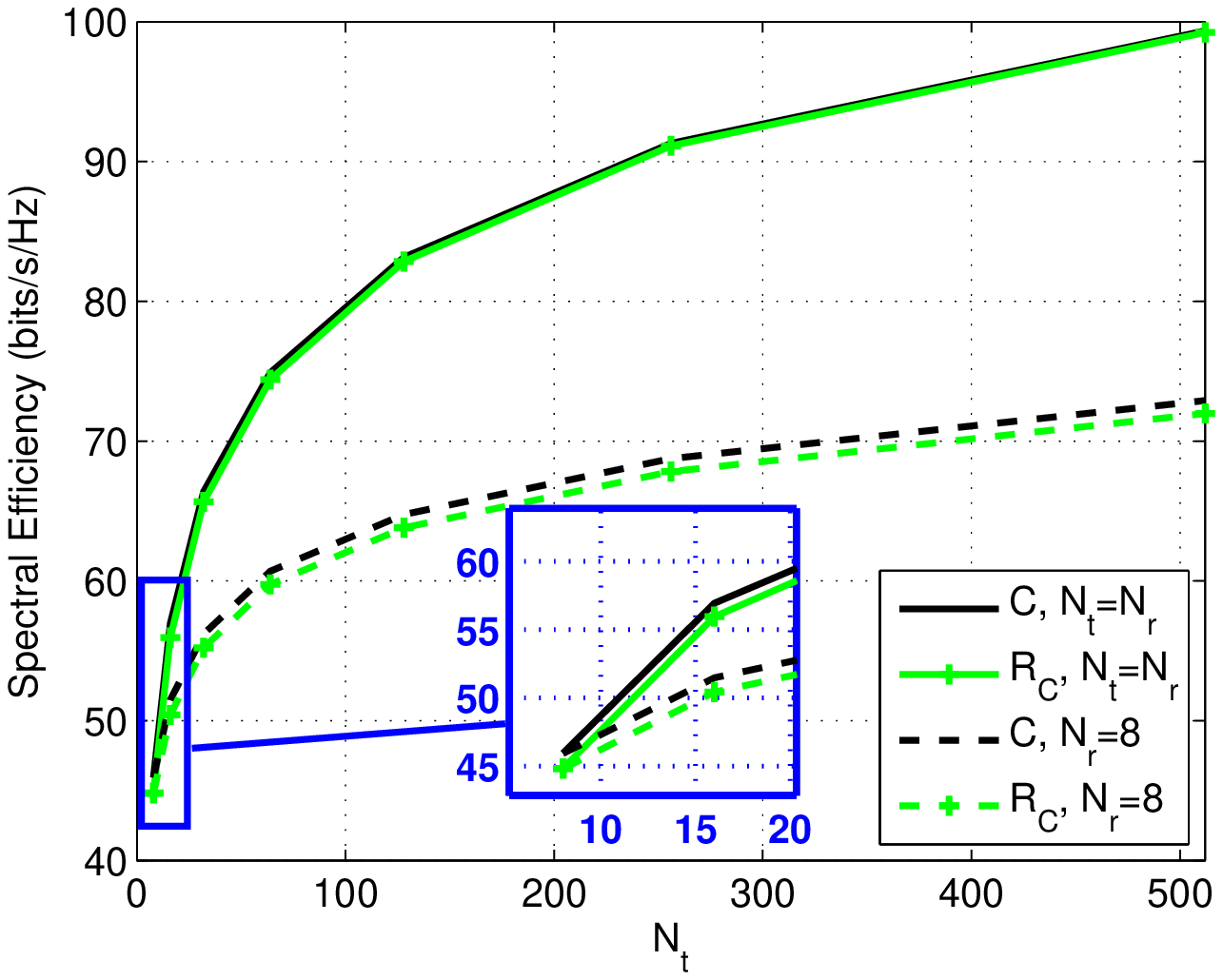}
\caption{$C, \: R_\text{C}, R_\text{C}^\text{A}$ when the number of the antennas varies, $\rho=34$ dB and geometry based channel with $L=5$.}
\label{fig:GeometryNumberofAntennas}
  \end{minipage}
\begin{minipage}[b]{0.48\textwidth}
 \includegraphics[width=3.2in]{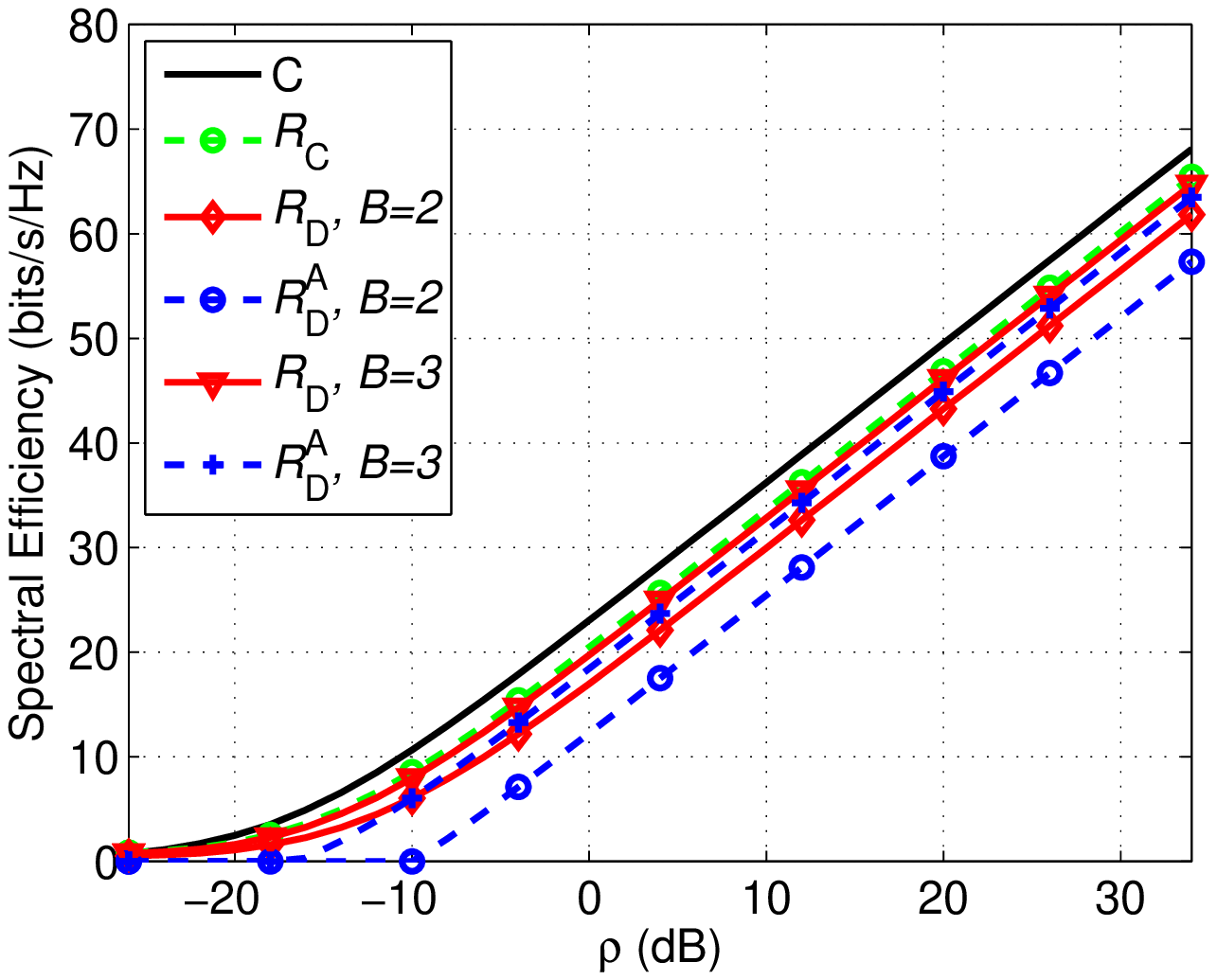}
\caption{Spectral efficiency achieved the hybrid beamformer with digital phase shifters based on (\ref{eq:SolutionDiscreteres}) $R_\text{D}$, compared to the bound based on Lemma 4 $R_\text{D}^\text{A}$, $R_\text{C}$ and $C$ for Rayleigh channel.}
\label{fig:RayleighRes}
  \end{minipage}
\hfill
  \begin{minipage}[b]{0.48\textwidth}
\includegraphics[width=3.2in]{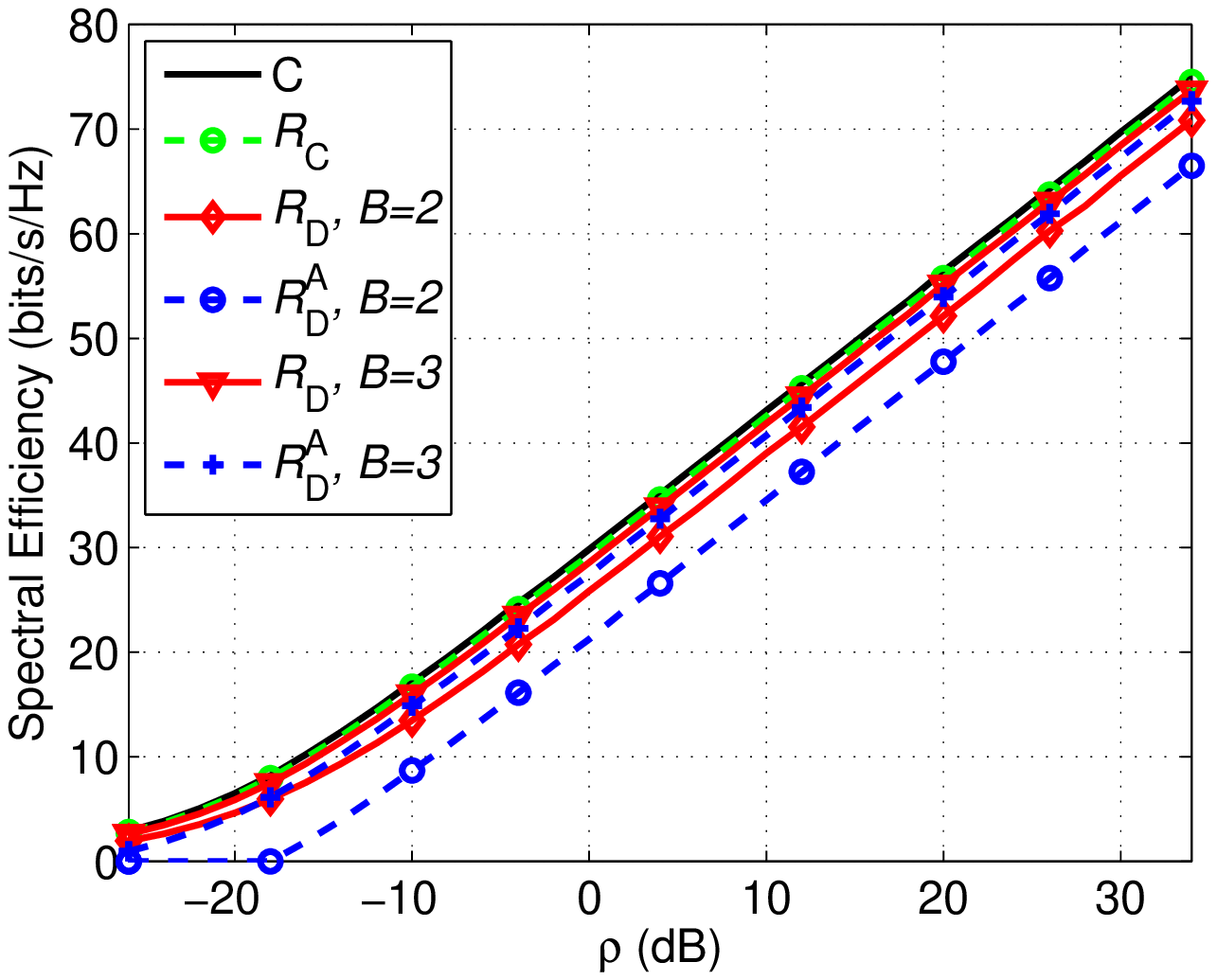}
\caption{Spectral efficiency achieved the hybrid beamformer with digital phase shifters based on (\ref{eq:SolutionDiscreteres}) $R_\text{D}$, compared to the bound based on Lemma 4 $R_\text{D}^\text{A}$, $R_\text{C}$ and $C$ for geometry based channel with $L=5$.}
\label{fig:GBPSRes}
  \end{minipage}
\end{figure}

In this section, the performance of the proposed hybrid beamforming schemes for the point-to-point and the multiuser scenarios operating in rich and sparse scattering channels is evaluated by Monte-Carlo simulations. The performance metric is average spectral efficiency over 1000 independent channel realizations and it is assumed that $M=K=4$. In this paper, all the closed-form expressions were derived for the scenario that $N_\text{t},N_\text{r} \to \infty$. To obtain the appropriate assumption on the number of the antenna elements for the simulations, the behavior of the hybrid beamformer with respect to $N_\text{t}$ and $N_\text{r}$ is first analyzed. In the following, the superscript \textquotedbl A\textquotedbl is used to distinguish the analytical results of the Lemmas 3, 4, 5 and 6 from the performance of the proposed schemes derived by the Monte-Carlo simulations. For example, the analytical spectral efficiency by the hybrid beamformer in Lemma 2 is expressed as $R_\text{C}^\text{A}=C-\Delta_\text{C}$ where $\Delta_\text{C}=\lim_{N_\text{t},N_\text{r} \to \infty} C-R_\text{C}$. The performance of the hybrid beamformer of Lemma 2 with the closed-form expression of Lemma 3 are investigated with respect to the number of antennas, where it is assumed that $N_\text{t}=N_\text{r}$ and then $N_\text{r}= 8$ for $N_\text{t} \in \lbrace 8,\: 16,\: 32,\: 64,\: 128,\: 256, \: 512 \rbrace$. Figure \ref{fig:RayleighNumberofAntennas} shows this performance $R_\text{C}^\text{A}$ compared to the simulation result, $R_\text{C}$, for the Rayleigh fading channel whereas Fig. \ref{fig:GeometryNumberofAntennas} presents these for the geometry based model. It is observed that $R_\text{C}^\text{A}$ and $R_\text{C}$ converge for both channels when $N_\text{t}=N_\text{r}$ is large. For the Rayleigh fading channel, $R_\text{C}^\text{A}$ predicts slightly lower spectral efficiency compared to the results from simulations $R_\text{C}$ when $N_\text{r}=8$ as shown in Fig. \ref{fig:RayleighNumberofAntennas}. On the other hand, it is observed from Fig. \ref{fig:GeometryNumberofAntennas} that $R_\text{C}^\text{A}$ is always larger than $R_\text{C}$ for the geometry based model as for this channel $C=R_\text{C}^\text{A}$.

Figure \ref{fig:RayleighRes} shows the performance of the hybrid beamformer with digital phase shifters, denoted as $R_\text{D}$, for a point-to-point system operating in rich scattering channel. It is observed that $R_\text{C}-R_\text{D}$ for $B=2$ and $B=3$ is 3.5 and 0.7 bits/s/Hz which is negligible compared to the high spectral efficiency achieved by large antenna arrays at high SNR. Hence, a simple rounding technique to set the discrete phases of the phase shifters with $B \geq 3$ can significantly simplify the calculations, and achieve a similar performance as analog phase shifters are employed. In addition, the lower-bound of the spectral efficiency based on Lemma 4, denoted as $R_\text{D}^\text{A}$, provides a good approximation when $B \geq 3$. For example, when $B=3$, $R_\text{D}-R_\text{D}^\text{A}$ is 1.2 bits/s/Hz. Figure \ref{fig:GBPSRes} presents a similar result for the sparse scattering channel. 
\begin{figure}
  \centering
  \begin{minipage}[b]{0.48\textwidth}
 \includegraphics[width=3.2in]{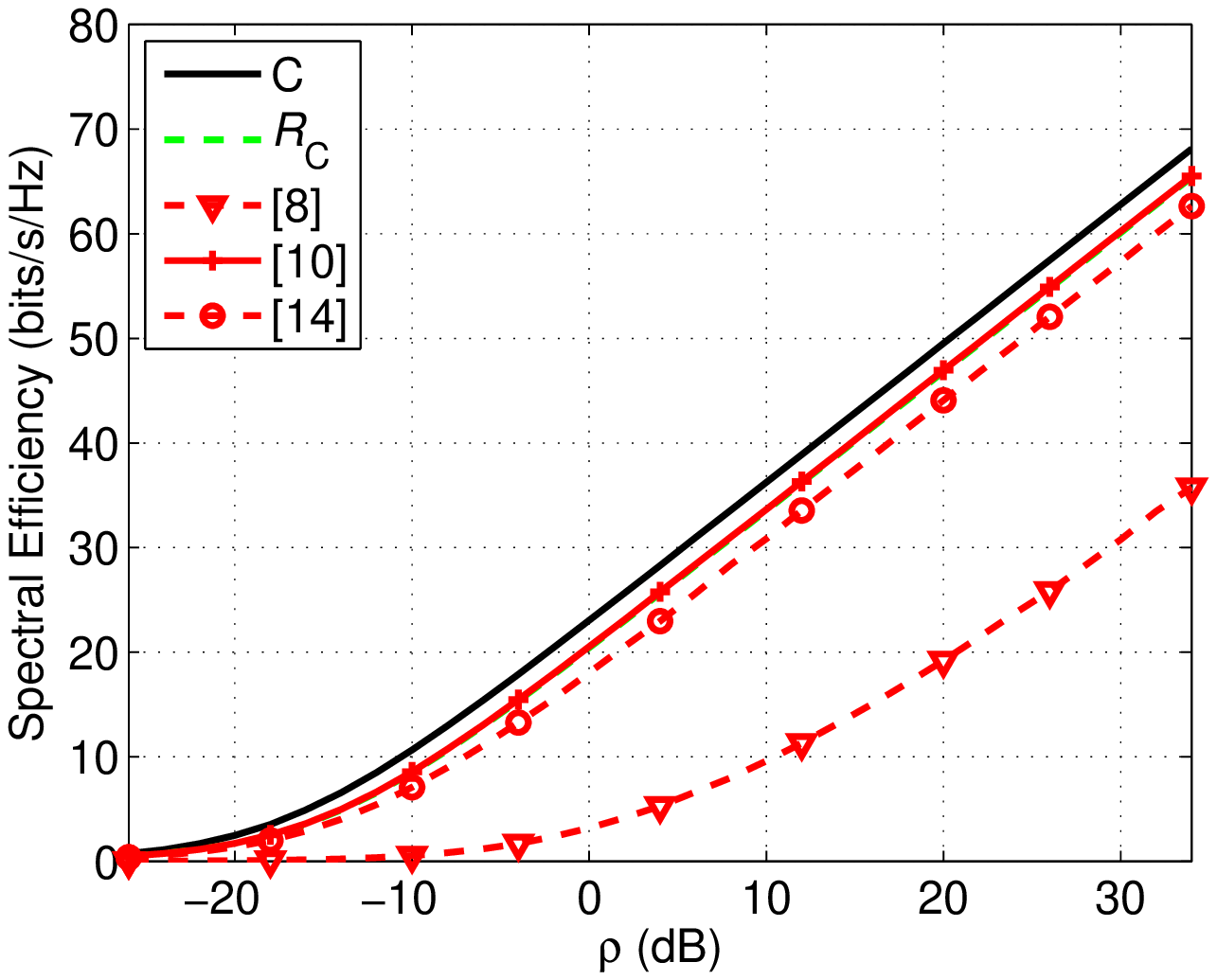}
\caption{Spectral efficiency achieved by the proposed algorithm compared to the state-of-the-art \cite{SpatiallySparsePrecodingAyach,NiDL15,FoadTorontoApril2015} when the wireless channel follows Rayleigh fading.}
\label{fig:ComparisonStateRayleigh}
  \end{minipage}
\hfill
\begin{minipage}[b]{0.48\textwidth}
\includegraphics[width=3.2in]{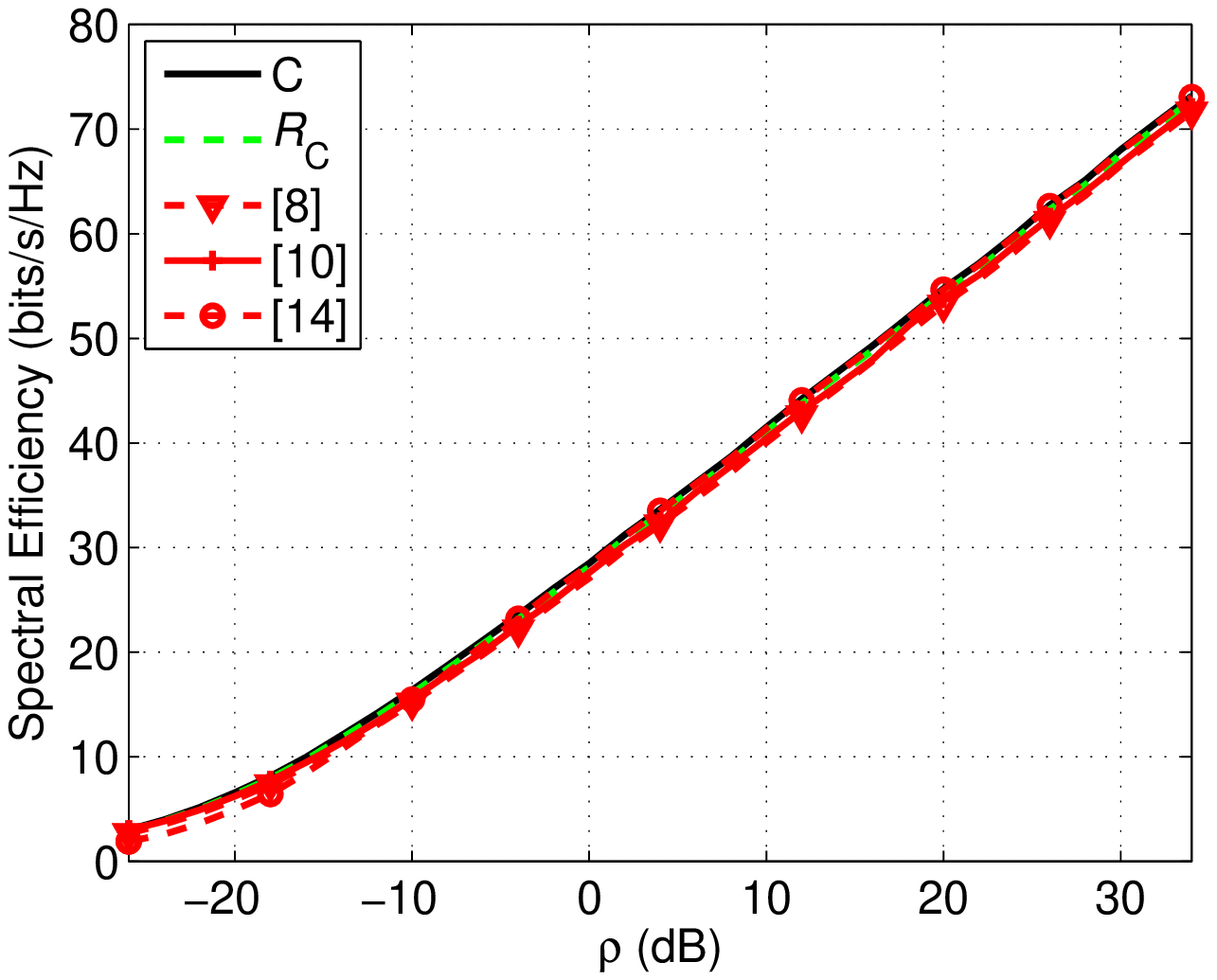}
\caption{Spectral efficiency achieved by the proposed algorithm compared to the state-of-the-art \cite{SpatiallySparsePrecodingAyach,NiDL15,FoadTorontoApril2015} when the wireless channel follows geometry based model with $L=5$.}
\label{fig:ComparisonStateSparse}
  \end{minipage}
  
  \begin{minipage}[b]{0.48\textwidth}
 \includegraphics[width=3.2in]{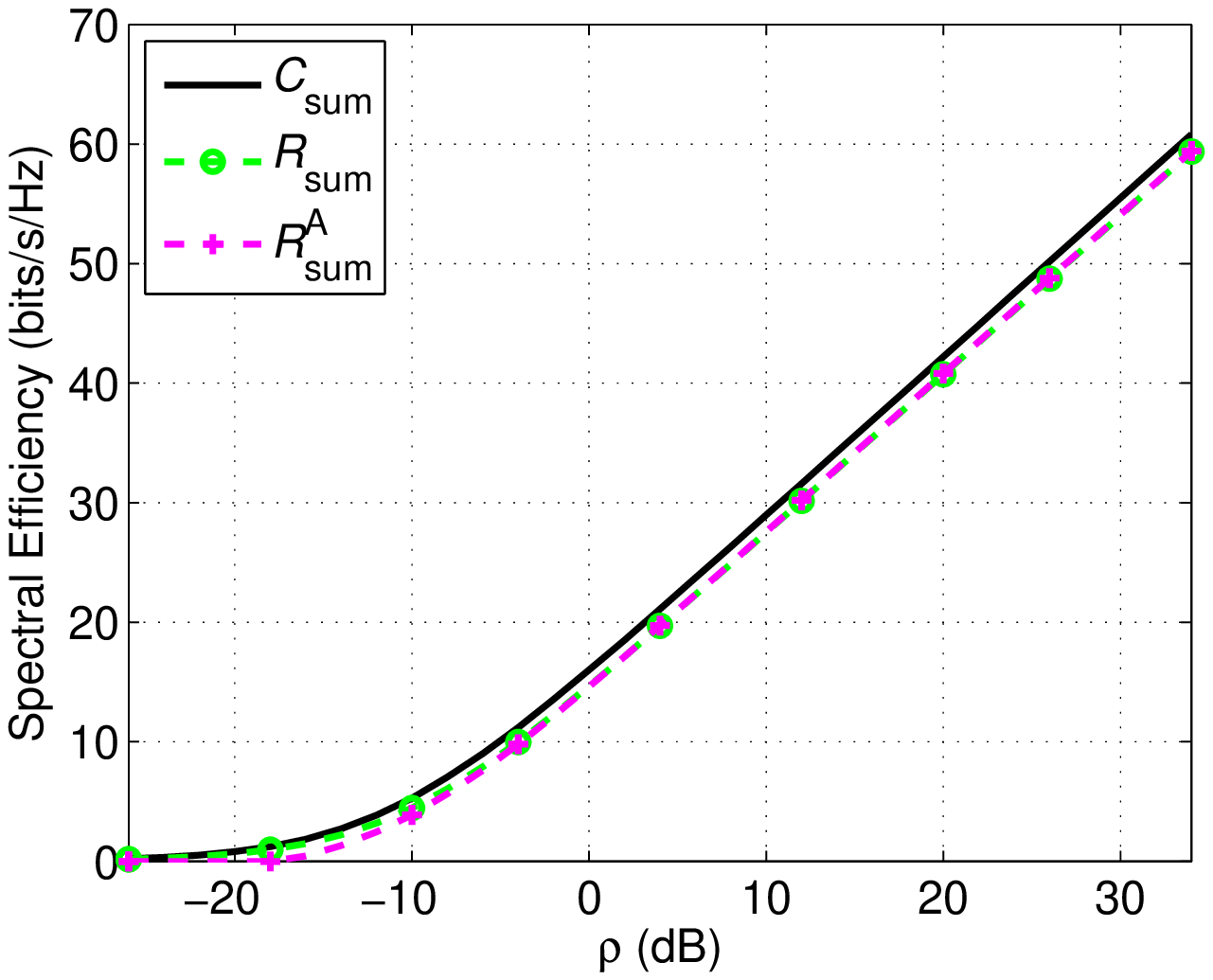}
\caption{Sum-rate achieved by ZF (digital beamforming) $C_\text{sum}$, the proposed hybrid beamformer for the multiuser scenario $R_\text{sum}$ and the bound based on Lemma 5 $R_\text{sum}^\text{A}$ for Rayleigh fading channel.}
\label{fig:ZF}
  \end{minipage}
\end{figure}
Figure \ref{fig:ComparisonStateRayleigh} and Fig. \ref{fig:ComparisonStateSparse} show the performance of the proposed algorithm in Lemma 2 compared to the state-of-the-art \cite{SpatiallySparsePrecodingAyach,NiDL15,FoadTorontoApril2015} for Rayleigh and geometry based channels. It is observed that the algorithm of \cite{SpatiallySparsePrecodingAyach} is not applicable to the Rayleigh fading channel, although it has a very good performance for the sparse scattering channel. The performance of the iterative algorithms of \cite{NiDL15} and \cite{FoadTorontoApril2015} is similar to the proposed scheme for both channels. 

For the downlink multiuser scenario with large number of antennas at the base station, ZF has been shown as the asymptotically optimal beamforming scheme in Rayleigh channels. Fig. \ref{fig:ZF} shows the achievable sum-rates by ZF with a digital beamformer and the proposed hybrid beamformer, denoted as $C_\text{sum}$ as $R_\text{sum}$, when $N_\text{t}=64$ and $K=4$. It is observed that the digital beamformer achieves 1.4 bit/s/Hz higher spectral efficiency than the hybrid beamformer as in Lemma 5.

\begin{figure}
  \centering
  \begin{minipage}[b]{0.48\textwidth}
 \includegraphics[width=3.2in]{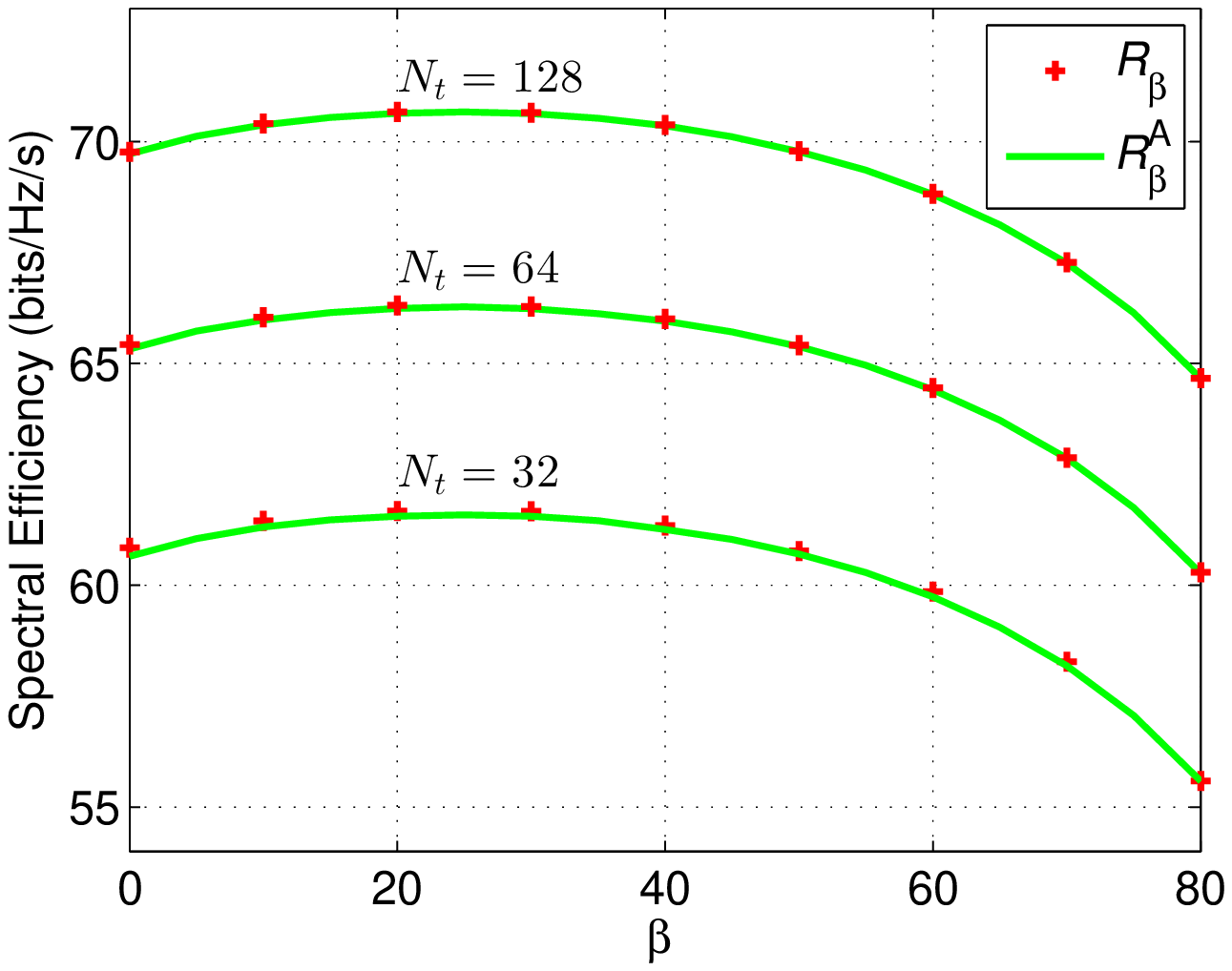}
\caption{Spectral efficiency achieved by the proposed phase shifter selection $R_\beta$, and the bound based on Lemma 6 $R_\beta^\text{A}$, $N_\text{t}=N_\text{r}$, $\rho=34$ dB.}
\label{fig:BetImpact}
  \end{minipage}
  \hfill
  \begin{minipage}[b]{0.48\textwidth}
\includegraphics[width=3.2in]{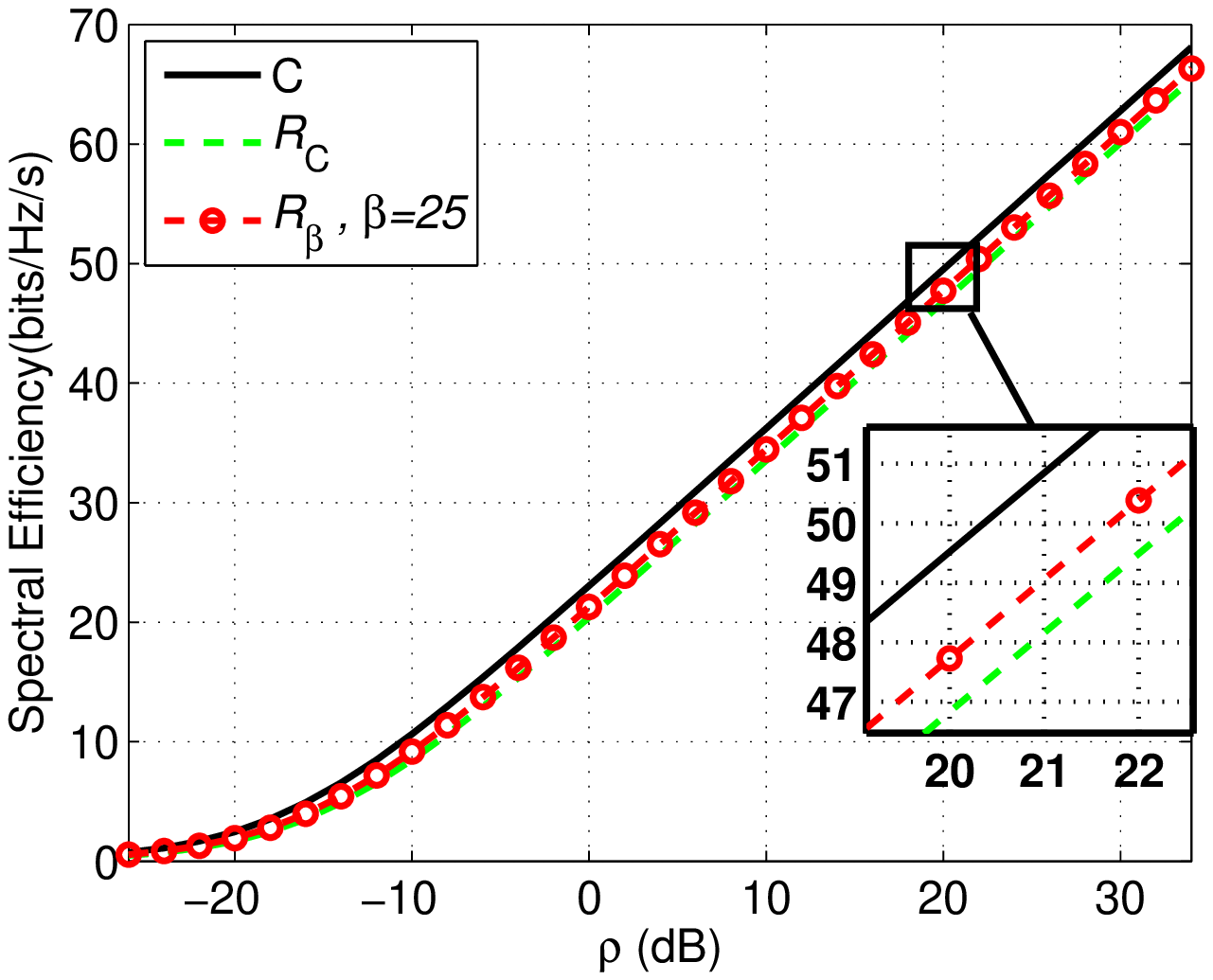}
\caption{Comparison of $C$, $R_\text{C}$ and the spectral efficiency achieved by the proposed phase shifter selection with $\beta = 25 $.}
\label{fig:BetaComparetoSVD}
  \end{minipage}
\end{figure}

Figure \ref{fig:BetImpact} shows the spectral efficiency achieved by the phase shifter selection scheme $R_\beta$ compared to the closed-form based on Lemma 6, denoted as $R_\beta^\text{A}$, for Rayleigh fading channel and different values of $\beta$ and $N_\text{t}$. It is observed that there is a good match between (\ref{eq:RProPSSelectionLemma}) and simulations. Compared to the scenario that all the phase shifters are in operation, the spectral efficiency can be improved when the phase shifter selection is applied with $0 <\beta < 50 $. In addition, the maximum performance is achieved when $\beta$ is around 25\%. Finally, Fig. \ref{fig:BetaComparetoSVD} presents the performance of the phase shifter selection scheme for $\beta =25 $ compared to $R_\text{C}$ and $C$. It is observed that $C-R_{\beta=25}$ is around $33 \%$ smaller than $C-R_\text{C}$. In addition, the spectral efficiency when all the phase shifters are in operation is almost equal to the case that $\beta\%=50\%$ of them are turned off which results in $50\%$ reduction in power consumption.

\section{Conclusion and Future Works}
\label{sec:conclusion}
In this paper, we derived the asymptotically optimal hybrid beamforming schemes to maximize the spectral efficiency for the point-to-point and multiuser systems with large antenna arrays, operating in rich and sparse scattering channels. The optimality of the solution was proved based on the properties of the singular vectors of the channel matrix. The elements of these vectors have a complex Gaussian distribution for Rayleigh fading model, and the singular vectors are equal to the steering vectors of the channel matrix for the geometry based model. In addition, we derived the closed-form expressions for the spectral efficiency when the proposed hybrid beamformer is used. It was shown that the performance of the hybrid beamformer, employing phase shifters with resolution more than 2-bits, can approach the performance of a similar system with analog phase shifters. In order to reduce the power consumption in the RF beamformer, a novel phase shifter selection scheme was proposed. This approach can increase the spectral efficiency and reduce the power consumption when channel follows Rayleigh fading model. Simulation results indicate that spectral efficiency improves when up to 50\% of the phase shifters are turned off. 

The hybrid beamformer investigated in this work was developed and evaluated under certain assumptions such as perfect CSI, narrowband systems, no RF impairments, ideal phase shifters and switches. However, in order to integrate hybrid beamformers into practical systems, the impact of these parameters should be investigated. 

\appendices
\section{Proof of Lemma 2}
\label{App:lemma2}
Defining the positive semidefinite matrices $\textbf{Q}, \: \tilde{\textbf{Q}}\in \mathbb{C}^{N_\text{t} \times N_\text{t}}$ as $\textbf{Q}=\frac{1}{N_\text{t}}\textbf{F}_\text{RF}  \textbf{F}_\text{B}  \textbf{P} \textbf{F}_\text{B}^\text{H} \textbf{F}_\text{RF}^\text{H} $ and $\tilde{\textbf{Q}}=\textbf{V}^\text{H}\textbf{Q} \textbf{V}$, the mutual information $I(\textbf{s},\textbf{y})$ is expressed as
\begin{align}
\label{eq:mutualRF1}
I(\textbf{s},\textbf{y})&= \log_2  \text{det}\Big (\textbf{I}_{N_\text{r}}+\rho \textbf{H} \textbf{Q} \textbf{H}^\text{H} \Big )=\log_2  \text{det} \Big ( \textbf{I}_{N_\text{r}}+ {\rho}  \textbf{U}\boldsymbol{\Sigma}\textbf{V}^\text{H}\textbf{QV }\boldsymbol{\Sigma}^\text{H}\textbf{U}^\text{H}  \Big ) \\ \nonumber
&=\log_2  \text{det} \Big ( \textbf{I}_{N_\text{r}}+ {\rho}  \boldsymbol{\Sigma}\tilde{\textbf{Q}}\boldsymbol{\Sigma}^\text{H} \Big )=\log_2  \text{det} \Big ( \textbf{I}_{N_\text{t}}+ {\rho} \boldsymbol{\Sigma}^\text{H} \boldsymbol{\Sigma}\tilde{\textbf{Q}}  \Big )  \\ \nonumber
& \stackrel{(b)}{\leq} \log_2 \Big( \prod_{{n_\text{t}}=1}^{N_\text{t}} (1+\rho  \bar{\sigma}_{n_\text{t}n_\text{t}}^2  \tilde{{Q}}_{n_\text{t}n_\text{t}}) \Big ),
\end{align}
where $\bar{\sigma}_{n_\text{t}n_\text{t}}^2 $ are the diagonal elements of $ \boldsymbol{\Sigma}^\text{H} \boldsymbol{\Sigma}$, and the inequality $(b)$ comes from linear algebra as for any positive semidefinite matrix $\textbf{A} \in \mathbb{C}^{N_\text{t} \times N_\text{t}}$, $ \text{det}( \textbf{A} ) \leq \prod_{n_\text{t}} A_{n_\text{t}n_\text{t}}$. If $\tilde{\textbf{Q}}$ is a diagonal matrix, then $(b)$ in (\ref{eq:mutualRF1}) turns into equality. Hence, the objective is to design $\textbf{F}_\text{RF}$ and $\textbf{F}_\text{B}$ such that they can diagonalize $\tilde{\textbf{Q}}$. In order to analyze $\tilde{\textbf{Q}}$, we investigate the behavior of the elements of $\textbf{G} \in \mathbb{C}^{N_\text{t} \times K}$, defiend as $\textbf{G}=1/\sqrt{N_\text{t}}\textbf{V}^\text{H}\textbf{F}_\text{RF}$, when $\frac{1}{\sqrt{N_\text{t}}}\textbf{v}_{n_\text{t}}^\text{H}\textbf{f}_{\text{RF},k}=0$ and $\frac{1}{\sqrt{N_\text{t}}}\textbf{v}_{n_\text{t}}^\text{H}\textbf{f}_{\text{RF},k}\neq 0$, $\forall \: n_\text{t}\neq k$. In the first case that $\frac{1}{\sqrt{N_\text{t}}}\textbf{v}_{n_\text{t}}^\text{H}\textbf{f}_{\text{RF},k}=0$ $\forall \: n_\text{t}\neq k'$, it could be easily shown that all of the elements of $\textbf{G}$ except the $G_{kk}$ become zero. Then, the last term in (\ref{eq:mutualRF1}) can be written as
\begin{align}
\label{eq:DiagQtilde}
\log_2 \Big( \prod_{{n_\text{t}}=1}^{N_\text{t}} (1+\rho  \sigma_{n_\text{t}n_\text{t}}^2  \tilde{{Q}}_{n_\text{t}n_\text{t}})\Big )
=\log_2 ( \prod_{k=1}^{K} (1+\rho  \sigma_{kk}^2  \tilde{{Q}}_{kk})\Big ).
\end{align}
On the other hand, if $\textbf{F}_\text{B}$ is a diagonal matrix, then $\textbf{F}_\text{B}  \textbf{P} \textbf{F}_\text{B}^\text{H} $ will have the same property. As a result, $\tilde{\textbf{Q}}=\textbf{G} \textbf{F}_\text{B}  \textbf{P} \textbf{F}_\text{B}^\text{H} \textbf{G}^\text{H}$ will also become a diagonal matrix since off-diagonal elements of \textbf{G} are zero. In addition, in (\ref{eq:AchievableRate4}), it was discussed that $\textbf{F}_\text{B}$ should be a unitary matrix to maximize the spectral efficiency. As $\textbf{F}_\text{B}$ is a diagonal and a unitary matrix, it could be concluded that $\vert F_{\text{B},kk} \vert^2 =1$. In this case, $\tilde{{Q}}_{kk}$ becomes
\begin{align}
\label{eq:optRFConti}
P_{kk} \vert G_{kk}\vert ^2=\frac{P_{kk}}{N_\text{t}} \vert \textbf{v}_{k}^\text{H} \textbf{f}_{\text{RF},k}\vert^2  = \frac{P_{kk}}{{N_\text{t}}}\vert \sum_{n_\text{t}=1}^{N_\text{t}} {V}^\ast_{n_\text{t}k} \text{e}^{j \theta_{n_\text{t}k}}  \vert ^2  \stackrel{(c)}{\leq}  \frac{P_{kk}}{{N_\text{t}}}\Big  \vert \sum_{n_\text{t}=1}^{N_\text{t}} \vert  V_{n_\text{t}k} \vert \Big \vert^2,
\end{align}
where the left hand side of $(c)$ is maximized when all the elements of $\textbf{v}_{k}$ are added constructively. In other words, $(c)$ in (\ref{eq:optRFConti}) turns into equality if $\textbf{F}_\text{RF}=\textbf{F}_\text{RF}^\text{opt1}$ as
\begin{align}
\label{eq:RFPreco1} 
F_{\text{RF},n_\text{t}k}^\text{opt1}=\text{e}^{j\angle V_{n_\text{t}k}}. 
\end{align}
In the following, we analyze the impact of setting $\textbf{F}_\text{RF}=\textbf{F}_\text{RF}^\text{opt1}$ on the off-diagonal elements of $\textbf{G}$ for rich and sparse scattering channels. For the Rayleigh channel, Theorem 1 expresses that the elements of singular vectors of the channel matrix are zero-mean i.i.d. random variables and their phases are uniformly distributed over $[0,\: 2\pi]$. As a consequence of law of large numbers
\begin{equation}
\label{eq:OffDiagRayleigh}
\lim_{N_\text{t} \to \infty}   \frac{1}{\sqrt{N_\text{t}}}\textbf{v}_{n_\text{t}}^\text{H}\textbf{f}_{\text{RF},k} =\lim_{N_\text{t} \to \infty}  \frac{1}{N_\text{t}} \sum \limits_{\substack{n_\text{t}'=1  }}^{N_\text{t}}\sqrt{N_\text{t}}  V_{n_\text{t}' n_\text{t}^{ }}^\ast \text{e}^{j \angle  V_{n_\text{t}'k}}  = \text{E}[ \sqrt{N_\text{t}} V_{n_\text{t}k}]=0,
\end{equation}
for $n_\text{t} \neq k$. For the geometry based model, Lemma 1 states that the RF precoder in (\ref{eq:RFPreco1}) becomes $\textbf{f}_{\text{RF},k}=\sqrt{N_\text{t}}\textbf{v}_k$, hence $1/\sqrt{N_\text{t}} \textbf{v}_{n_\text{t}}^\text{H}\textbf{f}_{\text{RF},k}= \textbf{v}_{n_\text{t} }^\text{H}\textbf{v}_{k}   =0, \: \forall \: n_\text{t} \neq k$. As a result, it could be concluded that all of the elements of \textbf{G} except the diagonal elements become zero for both channels, when $\textbf{F}_\text{RF}=\textbf{F}_\text{RF}^\text{opt1}$. As a result, the choice of $\textbf{F}_\text{RF}=\textbf{F}_\text{RF}^\text{opt1}$ and a diagonal $\textbf{F}_\text{B}$, with $\vert F_{\text{B},kk} \vert^2 =1$, imposes $(b)$ in (\ref{eq:mutualRF1}) to turn into equality. Finally, $I(\textbf{s},\textbf{y})$ is maximized when the diagonal matrix $\textbf{P}$ is calculated based on waterfilling.

It could be easily shown that when the hybrid beamformer at the receiver is also considered, by applying a similar RF beamformer at the receiver, $\sqrt{\frac{1}{N_\text{t}N_\text{r}}}\textbf{W}_\text{RF}^{\text{opt}^\text{H}}\textbf{H}\textbf{F}_\text{RF}^\text{opt}$ becomes a diagonal matrix for both channels. In addition, $\textbf{W}_\text{B}$ will have a similar structure to $\textbf{F}_\text{B}$. Hence, $\textbf{F}_\text{B}=\textbf{W}_\text{B}=\textbf{I}_K$, $\Gamma_\text{t}=N_\text{t}$, $\Gamma_\text{r}=N_\text{r}$ is the capacity achieving hybrid beamformer for both channels.   \hfill\(\Box\)

\section{Alternative Derivation of the RF Beamformer in Lemma 2}
\label{App:AlternativeProof}

Since $\textbf{V}$ is a unitary matrix, $\Vert  \sqrt{\frac{1}{N_\text{t}}} \textbf{F}_{\text{RF}}- \textbf{F}_\text{d} \Vert^2=\Vert \sqrt{\frac{1}{N_\text{t}}} \textbf{V}^\text{H} \textbf{F}_{\text{RF}}- \textbf{V}^\text{H} \textbf{F}_\text{d} \Vert^2$. It could be easily verified that
\begin{equation}
\label{eq:alternativeProodHelp}
 \Vert \sqrt{\frac{1}{N_\text{t}}} \textbf{V}^\text{H} \textbf{F}_{\text{RF}}- \textbf{V}^\text{H} \textbf{F}_\text{d} \Vert^2 \geq \Vert \sqrt{\frac{1}{N_\text{t}}} \textbf{F}_\text{d}^\text{H} \textbf{F}_{\text{RF}}- \textbf{I}_K\Vert^2 , \: \: \: \: \text{s.t.  } \vert F_{\text{RF},n_\text{t}k}\vert^2=1 .
\end{equation}
The right hand side of the inequality can be reformulated as
\begin{align}
\label{eq:RFOpt2}
&\underset{\textbf{F}_{\text{RF}}}{\text{minimize}} \quad \sum_{k=1}^K \Big \vert \sqrt{\frac{1}{N_\text{t}}} \textbf{f}_{\text{d},k}^\text{H} \textbf{f}_{\text{RF},k}-1 \Big\vert ^2+\sum_{k'=1}^K \sum\limits_{\substack{k=1 \\ k \neq k' }}^K  \Big \vert \sqrt{\frac{1}{N_\text{t}}} \textbf{f}_{\text{d},k}^\text{H} \textbf{f}_{\text{RF},k'} \Big \vert ^2, \: \: \: \: \text{s.t.  } \vert F_{\text{RF},n_\text{t}k}\vert^2=1 .
\end{align}
The cost function can be lower-bounded as
\begin{align}
\label{eq:RFOpt3}
&\min \Big ( \sum_{k=1}^K \Big \vert \sqrt{\frac{1}{N_\text{t}}}\textbf{f}_{\text{d},k}^\text{H} \textbf{f}_{\text{RF},k}-1 \Big \vert ^2+\sum_{k'=1}^K \sum\limits_{\substack{k=1 \\ k \neq k' }}^K  \Big\vert \sqrt{\frac{1}{N_\text{t}}} \textbf{f}_{\text{d},k}^\text{H} \textbf{f}_{\text{RF},k'} \Big\vert ^2 \Big) \\ \nonumber
& = \min \Big ( \sum_{k=1}^K \Big \vert \sqrt{\frac{1}{N_\text{t}}} \textbf{f}_{\text{d},k}^\text{H} \textbf{f}_{\text{RF},k}-1 \Big\vert ^2 \Big)+ \min \Big(\sum_{k'=1}^K \sum\limits_{\substack{k=1 \\ k \neq k' }}^K  \Big \vert \sqrt{\frac{1}{N_\text{t}}}  \textbf{f}_{\text{d},k}^\text{H} \textbf{f}_{\text{RF},k'} \Big \vert ^2 \Big) \\ \nonumber
& \stackrel{(d)}{\geq}\min \Bigg ( \sum_{k=1}^K \Big ( \Big \vert \sqrt{\frac{1}{N_\text{t}}} \textbf{f}_{\text{d},k}^\text{H} \textbf{f}_{\text{RF},k} \Big \vert  -1 \Big)^2 \Bigg) +\min \Big(\sum_{k'=1}^K \sum\limits_{\substack{k=1 \\ k \neq k' }}^K \Big \vert \sqrt{\frac{1}{N_\text{t}}} \textbf{f}_{\text{d},k}^\text{H} \textbf{f}_{\text{RF},k'} \Big \vert ^2 \Big) \\ \nonumber
&  \stackrel{(e)}{\geq}    \sum_{k=1}^K \min \Big (\Big \vert \sqrt{\frac{1}{N_\text{t}}} \textbf{f}_{\text{d},k}^\text{H} \textbf{f}_{\text{RF},k} \Big \vert  -1 \Big)^2 \stackrel{(f)}{=}    \sum_{k=1}^K  \Bigg (\max \Big ( \Big \vert\sqrt{\frac{1}{N_\text{t}}} \textbf{f}_{\text{d},k}^\text{H} \textbf{f}_{\text{RF},k}\Big \vert \Big)  -1 \Bigg)^2 ,
\end{align}
where $(f)$ comes from the fact that $ \Big \vert \sqrt{\frac{1}{N_\text{t}}} \textbf{f}_{\text{d},k}^\text{H} \textbf{f}_{\text{RF},k} \Big \vert \leq 1$. Hence, the last term in (\ref{eq:RFOpt3}) is minimized if
\begin{align}
\label{eq:RFOptSimplified}
&\underset{\textbf{f}_{\text{RF},k}}{\text{maximize}} \quad \Big \vert \sqrt{\frac{1}{N_\text{t}}} \textbf{f}_{\text{d},k}^\text{H}\textbf{f}_{\text{RF},k} \Big \vert , \: \: \: \: \text{s.t.  } \vert F_{\text{RF},n_\text{t}k}\vert^2=1 ,
\end{align}
which is similar to (\ref{eq:optRFConti}) in Appendix (\ref{App:lemma2}). It was shown that $\vert \sqrt{\frac{1}{N_\text{t}}}\textbf{f}_{\text{d},k}^\text{H} \textbf{f}_{\text{RF},k'}\vert=0 $, $\forall k \neq k'$ and $\sqrt{\frac{1}{N_\text{t}}}\textbf{f}_{\text{d},k}^\text{H}\textbf{f}_{\text{RF},k}$ becomes a real and positive number when $F_{\text{RF},n_\text{t}k}^\text{opt}=\text{e}^{j\angle F_{\text{d},n_\text{t}k}}.$ Hence, $(e)$ and $(d)$ turn into equality, and the cost function in (\ref{eq:RFOpt2}) is minimized. Finally, (\ref{eq:alternativeProodHelp}) turns into equality and $\Vert  \sqrt{\frac{1}{N_\text{t}}} \textbf{F}_{\text{RF}}- \textbf{F}_\text{d} \Vert^2$ is minimized.\hfill\(\Box\)
\section{Proof of of Lemma 3}
\label{App:Lemma3Proof}
As a result of Theorem 1, E$[\sqrt{N_\text{t}} \vert V_{n_\text{t}k} \vert]=$E$[\sqrt{N_\text{r}} \vert U_{n_\text{r}k} \vert] =\frac{\sqrt{\pi}}{2}$, and hence,
\begin{align}
\label{eq:DiagRayleigh}
\lim_{N_\text{t} \to \infty} \frac{1}{\sqrt{N_\text{t}}}\vert \textbf{v}_\text{k}^\text{H} \textbf{f}_{\text{RF},k}^\text{opt} \vert &=\lim_{N_\text{r} \to \infty}\frac{1}{\sqrt{N_\text{r}}}\vert \textbf{u}_\text{k}^\text{H} \textbf{w}_{\text{RF},k}^\text{opt} \vert=\lim_{N_\text{t} \to \infty}\frac{1}{{N_\text{t}}}\Big  \vert \sum_{n_\text{t}=1}^{N_\text{t}}  \vert\sqrt{N_\text{t}} V_{n_\text{t}k} \vert \Big \vert =\frac{\sqrt{\pi}}{2}.
\end{align}  
Referring to the matrix $\textbf{G}=1/\sqrt{N_\text{t}} \textbf{V}^\text{H}\textbf{F}_\text{RF}$ in Appendix \ref{App:lemma2}, $G_{kk}=\sqrt{\pi}/2$ and $G_{n_{\text{t}}k}=0, \: \forall n_\text{t}\neq k$. Applying a similar RF beamformer at the receiver side, it could be easily verified that $\textbf{R}_\text{n}=1/N_\text{r} \textbf{W}_\text{RF}^\text{H}\textbf{W}_\text{RF}=\textbf{I}_K$. The spectral efficiency in (\ref{eq:AchievableRate1}) at high SNR becomes 
\begin{align}
\label{eq:RProposed}
R_\text{C} & =\lim_{N_\text{t}\to \infty} \lim_{N_\text{r}\to \infty}  \log_2  \text{det} \Big ( \frac{\rho}{N_\text{t}N_\text{r}}  \textbf{R}_\text{n}^{-1}\textbf{W}_\text{B}^\text{H}\textbf{W}_\text{RF}^\text{H}\textbf{H} \textbf{F}_\text{RF} \textbf{F}_\text{B} \textbf{P}\textbf{F}_\text{B}^\text{H}\textbf{F}_\text{RF}^\text{H} \textbf{H}^\text{H}\textbf{W}_\text{RF}\textbf{W}_\text{B}\Big ) \\ \nonumber
&= \lim_{N_\text{t}\to \infty} \lim_{N_\text{r}\to \infty} \log_2 \text{det} \Big ( \frac{\rho}{N_\text{t}N_\text{r}}  \textbf{W}_\text{RF}^\text{H}\textbf{U} \boldsymbol{\Sigma} \textbf{V}^\text{H} \textbf{F}_\text{RF}  \textbf{P}\textbf{F}_\text{RF}^\text{H} \textbf{V} \boldsymbol{\Sigma} \textbf{U}^\text{H}\textbf{W}_\text{RF}\Big ) \\ \nonumber
&= \log_2 \text{det}\Bigg ( \Big ( \frac{\pi}{4}\Big ) ^2 \rho  \boldsymbol{\Sigma'}^2   \textbf{P}\Bigg  )=\sum_{k=1}^K \log_2 (\frac{\pi^2}{4^2} \rho  P_{kk} \sigma_{kk}^2)\\ \nonumber
&=\sum_{k=1}^K \log_2 ( \rho  P_{kk} \sigma_{kk}^2)+2K\log_2 (\frac{\pi}{4}),
\end{align}
where $\boldsymbol{\Sigma'}=$diag$(\sigma_1^2, \: ..., \: \sigma_K^2)$. Considering that the first term in the last line is $C$ in (\ref{eq:Hcapacity}), the lemma is proved. \hfill\(\Box\)
\section{Proof of OF Lemma 4}
\label{App:DisAlg2}
In Appendix \ref{App:lemma2}, it was shown that the achievable rate depends on $\frac{1}{\sqrt{N_\text{t}}}\vert \textbf{v}_{k}^\text{H} \textbf{f}_{\text{RF},k} \vert$. Letting $\delta_{n_\text{t}k}=\theta_{n_\text{t}k}^\text{d}-\angle F_{\text{d},n_\text{t}k}$ where $\frac{-2 \pi}{2^{B+1}} \leq \delta_{n_\text{t}k} \leq \frac{2 \pi}{2^{B+1}}$, 
\begin{align}
\frac{1}{\sqrt{N_\text{t}}}\vert \textbf{v}_{k}^\text{H} \textbf{f}_{\text{RF},k} \vert &=\frac{1}{\sqrt{N_\text{t}}} \Big \vert \sum_{n_\text{t}=1}^{N_\text{t}} \vert V_{n_\text{t}k}\vert \text{e}^{-j \angle V_{n_\text{t}k}} \text{e}^{j \theta_{n_\text{t}k}^{\ast}} \Big \vert =\frac{1}{\sqrt{N_\text{t}}} \Big \vert \sum_{n_\text{t}=1}^{N_\text{t}} \vert V_{n_\text{t}k}\vert \text{e}^{j\delta_{n_\text{t}k}} \Big \vert \\ \nonumber
&=\frac{1}{\sqrt{N_\text{t}}} \Bigg \vert \sum_{n_\text{t}=1}^{N_\text{t}} \vert V_{n_\text{t}k}\vert \Big (\cos(\delta_{n_\text{t}k})+j \sin (\delta_{n_\text{t}k}) \Big) \Bigg\vert \geq \frac{1}{\sqrt{N_\text{t}}} \Big \vert \sum_{n_\text{t}=1}^{N_\text{t}} \vert V_{n_\text{t}k}\vert \cos(\delta_{n_\text{t}k}) \Big \vert \\ \nonumber 
&\geq  \frac{1}{\sqrt{N_\text{t}}} \cos(\frac{2 \pi}{2^{B+1}}) \sum_{n_\text{t}=1}^{N_\text{t}} \vert V_{n_\text{t}k}\vert .
\end{align}
It could be easily shown that $1/\sqrt{N_\text{t}}\vert \textbf{v}_k^\text{H}\textbf{f}_{k'} \vert=0\: \forall k \neq k'$ holds for both channel models. Following a similar approach as in Appendix \ref{App:Lemma3Proof}, the rest of the proof is straight forward. \hfill\(\Box\)
\section{Proof of Lemma 5}
\label{AppendixMU}
Letting $\textbf{H}=\textbf{U}\boldsymbol{\Sigma} \textbf{V}^\text{H}$ and $\textbf{F}_\text{d}=\textbf{V}_{1:K}$, the sum-rate capacity of a multiuser broadcast channel can be expressed as \cite{WishvanathTse}
\begin{align}
C_\text{sum}(P_\text{t},\textbf{H}) &= \max_{\text{trace(\textbf{P})}\leq 1}\log_2 \text{det} \Big (\textbf{I}_K+\frac{P_\text{t}}{\sigma_z^2}\textbf{P} \textbf{H}\textbf{H}^\text{H} \Big) =
\max_{\text{trace(\textbf{P})}\leq 1}\log_2 \text{det} \Big (\textbf{I}_K+\frac{P_\text{t}}{\sigma_z^2}\textbf{P} \textbf{U}\boldsymbol{\Sigma} \textbf{V}^\text{H} \textbf{V}\boldsymbol{\Sigma}^\text{H}\textbf{U}^\text{H}  \Big)  \\ \nonumber
&=\max_{\text{trace(\textbf{P})}\leq 1}\log_2 \text{det} \Big (\textbf{I}_K+\frac{P_\text{t}}{\sigma_z^2}\textbf{P} \textbf{U}\boldsymbol{\Sigma}_{1:K}^\text{2}\textbf{U}^\text{H}  \Big)=C_\text{sum}( P_\text{t},\textbf{H}\textbf{F}_\text{d}),
\end{align}
where the last two equalities comes from the fact that $\boldsymbol{\Sigma}$ has only $K$ nonzero elements and $\boldsymbol{\Sigma}_{1:K}^\text{2} =\boldsymbol{\Sigma} \textbf{V}^\text{H} \textbf{V}\boldsymbol{\Sigma}^\text{H}=\boldsymbol{\Sigma}\textbf{V}^\text{H}\textbf{F}_\text{d} \textbf{F}_\text{d}^\text{H} \textbf{V}\boldsymbol{\Sigma}^\text{H}$. As the singular vectors of the channel are in the direction of the channel steering vectors, it can be easily concluded that the RF beamformer of Lemma 2 is virtually the optimal scheme for the sparse channel model. 

For the Rayleigh channel employing $\textbf{F}_\text{d}$ as the RF beamformer is equivalent to relaxing the constant modulus constraint of the phase shifters. When $N_\text{t}\to \infty$, the performance of ZF beamformer with $\textbf{F}_\text{ZF}=\textbf{H}^\text{H}(\textbf{HH}^\text{H})^{-1}$ converges to the sum-capacity \cite{MasiveMIMORusek}, and the channel input-output relationship becomes 
\begin{equation}
\textbf{y}=\sqrt{\frac{P_\text{t}}{\Gamma_\text{t}}} \textbf{HH}^\text{H}(\textbf{HH}^\text{H})^{-1} \textbf{s}+\textbf{z}=\sqrt{\frac{P_\text{t}}{\Gamma_\text{t}}} \textbf{s}+\textbf{z}.
\end{equation}
In this case, $\Gamma_\text{t}$ in (\ref{eq:MUGamma}) is 
\begin{align}
\Gamma_\text{t}&=\text{E}\Big[ \frac{1}{K}\text{trace}  (\textbf{F}_\text{ZF}\textbf{F}_\text{ZF}^\text{H}  )\Big]  =\frac{1}{K}\text{E}\Bigg[\text{trace}  \bigg (\textbf{H}^\text{H} (\textbf{HH}^\text{H} )^{-1} \big [ (\textbf{HH}^\text{H} )^{-1}\big]^\text{H}\textbf{H}  \bigg)\Bigg] \\ \nonumber
&=\frac{1}{K}\text{E}\Bigg[\text{trace}   \bigg (\textbf{H}\textbf{H}^\text{H} (\textbf{HH}^\text{H} )^{-1} \big [ (\textbf{HH}^\text{H} )^{-1}\big]^\text{H}  \bigg)\Bigg]= \frac{1}{K} \text{E}\Bigg[\text{trace}\Big( (\textbf{HH}^\text{H})^{-1} \Big) \Bigg]\\ \nonumber 
&=\frac{1}{N-K},
\end{align}
as $\text{E}[\text{trace}  \big( (\textbf{HH}^\text{H} )^{-1} \big)=K/(N-K)$ for central complex Wishart matrices \cite{TulinoMatrix}. The spectral efficiency achieved by ZF is expressed as 
\begin{align}
C_\text{sum}(P_\text{t},\textbf{H})&=C_\text{sum}(P_\text{t},\frac{1}{\sqrt{\Gamma_\text{t}}}\textbf{H}\textbf{F}_\text{ZF})= K\log_2 (1+\rho) \\ \nonumber 
&=K\log_2 (1+\frac{P_\text{t}\text{E}[\vert s_k\vert^2]}{\Gamma_\text{t}\sigma_z^2}) =K\log_2 (1+\frac{P_\text{t}}{K\Gamma_\text{t} \sigma_z^2}),
\end{align}
where E$[\textbf{ss}^\text{H}]=1/K\textbf{I}_{K}$ and $\rho$ is the received SNR at the user side.

In addition, by applying ZF to the effective channel $\textbf{H}_\text{e}=\textbf{H}\textbf{F}_\text{d}=\textbf{U}\boldsymbol{\Sigma}_{1:K}$, the precoder matrix becomes $\textbf{F}_\text{ZFe}=\textbf{H}_\text{e}^{-1}=\boldsymbol{\Sigma}_{1:K}^{-1}\textbf{U}^\text{H}$. It should be noted that $\textbf{F}_\text{d}^\text{H}\textbf{F}_\text{d}=\textbf{I}_K$, and the rank of $\textbf{H} \in \mathbb{C}^{K \times N_\text{t}}$ is $K$ and hence $\boldsymbol{\Sigma}$ has only $K$ nonzero elements. Then, the normalization factor $\Gamma_\text{t}$ can be calculated as 
\begin{align}
\Gamma_\text{t}&=\frac{1}{K}\text{E}\Big[\text{trace}(\textbf{F}_\text{d} \textbf{F}_\text{ZFe} \textbf{F}_\text{ZFe} ^\text{H}\textbf{F}_\text{d} ^\text{H})\Big]=\frac{1}{K}\text{E}\Big[\text{trace}(\textbf{F}_\text{ZFe} \textbf{F}_\text{ZFe} ^\text{H})\Big] =\frac{1}{K}\text{E}\Big[ \text{trace}(\boldsymbol{\Sigma}_{1:K}^{-1}\textbf{U}^\text{H}  \textbf{U}\boldsymbol{\Sigma}_{1:K}^{-1} )\Big]\\ \nonumber
&=\frac{1}{K}\text{E}\Big[\text{trace}(\boldsymbol{\Sigma}_{1:K}^{-2} )\Big]= \frac{1}{K}\text{E}\Big[\text{trace}\big( (\boldsymbol{\Sigma}\boldsymbol{\Sigma}^{\text{H}})^{-1} \big )\Big]=\frac{1}{K}\text{E}\Big[\text{trace}\big( (\boldsymbol{\Sigma}\textbf{V}^\text{H}\textbf{V}\boldsymbol{\Sigma}^{\text{H}})^{-1} \textbf{U}^\text{H} \textbf{U} \big )\\ \nonumber
&=\frac{1}{K}\text{E}\Big[ \text{trace}\big( (  \textbf{U} \boldsymbol{\Sigma}\textbf{V}^\text{H}\textbf{V}\boldsymbol{\Sigma}^{\text{H}} \textbf{U}^\text{H} )^{-1} \big )\Big]=\frac{1}{K} \text{E}\Big[ \text{trace}\Big( (\textbf{HH}^\text{H})^{-1} \Big)\Big].
\end{align}
As a consequence, $1/\sqrt{\Gamma_\text{t}}\textbf{H}\textbf{F}_\text{ZF}=1/\sqrt{\Gamma_\text{t}}\textbf{H}\textbf{F}_\text{d}\textbf{F}_\text{ZFe}$ and
\begin{equation} 
\label{eq:ZFHe}
C_\text{sum}(P_\text{t},\textbf{H})=C_\text{sum}(P_\text{t},\frac{1}{\sqrt{\Gamma_\text{t}}}\textbf{H}_\text{e}  \textbf{F}_\text{ZFe}).
\end{equation} 
Hence, the asymptotically optimal hybrid beamforming scheme is derived when the constant modulus constraint at the RF beamformer is relaxed.

Since $K$ is fixed and $N_\text{t}\to \infty$, the array gain and therefore the received SNR grow large. Hence, the asymptotic behavior of MIMO channels at high SNR can be applied. In Theorem 3 of \cite{1207369} and Theorem 2 of \cite{HighSNRjindal}, it was shown that
\begin{equation}
\lim_{\rho \to \infty} \big[ C(P_\text{t},\textbf{H})-C_\text{sum}(P_\text{t},\textbf{H}) \big]=0, 
\end{equation}
where $ C(P_\text{t},\textbf{H})$ is the capacity of the point-to-point system. Considering $C(P_\text{t},\textbf{H})=C(P_\text{t},\textbf{U}^\text{H}\textbf{H})$, it could be concluded that 
\begin{equation}
\label{eq:rotations}
C_\text{sum}(P_\text{t},\textbf{U}^\text{H}\textbf{H})=C(P_\text{t},\textbf{U}^\text{H}\textbf{H})=C(P_\text{t},\textbf{H})=C_\text{sum}(P_\text{t},\textbf{H}). 
\end{equation}
Let $R_\text{sum}(P_\text{t},\frac{1}{\sqrt{N_\text{t}}} \textbf{H}\textbf{F}_\text{RF})$ denote the achievable sum-rate of multiuser scenario when the constant modulus is taken into account. Similar to (\ref{eq:rotations}), it could be easily verified that 
\begin{align}
\label{eq:MUFEquivalent1}
R_\text{sum}(P_\text{t},\frac{1}{\sqrt{N_\text{t}}} \textbf{H}\textbf{F}_\text{RF})&=  R_\text{sum}(P_\text{t},\frac{1}{\sqrt{N_\text{t}}} \textbf{U}^\text{H} \textbf{H}\textbf{F}_\text{RF})\\ \nonumber
&=\max_{\text{trace(\textbf{P})}\leq 1}\log_2 \text{det} \Big (\textbf{I}_K+\frac{P_\text{t}}{N_\text{t} \sigma_z^2}\textbf{P}\boldsymbol{\Sigma}\textbf{V}^\text{H}\textbf{F}_\text{RF}  \textbf{F}_\text{RF}^\text{H} \textbf{V} \boldsymbol{\Sigma}^\text{H}  \Big).
\end{align}
Now, the RF beamformer that maximizes $R_\text{sum}(P_\text{t},\frac{1}{\sqrt{N_\text{t}}} \textbf{U}^\text{H} \textbf{H}\textbf{F}_\text{RF})$ is obtained by 
\begin{align}
\textbf{F}_\text{RF}^{\text{opt}}=\underset{\textbf{F}_\text{RF}}{\arg\max}\: R_\text{sum}(P_\text{t},\frac{1}{\sqrt{N_\text{t}}}  \textbf{U}^\text{H}\textbf{H}\textbf{F}_\text{RF}), \text{    s.t. } \vert F_{\text{RF},n_\text{t}k} \vert =1.
\end{align}
Similar to Appendix \ref{App:lemma2}, $\textbf{F}_\text{RF}^{\text{opt}}$ of Lemma 2 that can diagonalize $\textbf{P}\boldsymbol{\Sigma}\textbf{V}^\text{H}\textbf{F}_\text{RF}  \textbf{F}_\text{RF}^\text{H} \textbf{V} \boldsymbol{\Sigma}^\text{H} $ will also maximize $R_\text{sum}(P_\text{t},\frac{1}{\sqrt{N_\text{t}}}  \textbf{U}^\text{H}\textbf{H}\textbf{F}_\text{RF})$ in (\ref{eq:MUFEquivalent1}). On the other hand, in Appendix C it is shown that $\frac{1}{\sqrt{N_\text{t}}} \textbf{H}\textbf{F}_\text{RF}^{\text{opt}}=\frac{\sqrt{\pi}}{2}\textbf{U} \boldsymbol{\Sigma}\textbf{V}^\text{H} \textbf{F}_\text{d}=\frac{\sqrt{\pi}}{2}\textbf{H}_\text{e}$. Additionally, in (\ref{eq:ZFHe}) it was discussed that $\textbf{F}_\text{ZFe}$ is asymptotically optimal for $\textbf{H}_\text{e}$. As a result,
\begin{align}
R_\text{sum}(P_\text{t},\frac{1}{\sqrt{N_\text{t}}} \textbf{U}^\text{H} \textbf{H}\textbf{F}_\text{RF}^{\text{opt}})&=
 \max_{\text{trace(\textbf{P})}\leq 1}\log_2 \text{det} \Big (\textbf{I}_K+\frac{P_\text{t}}{N_\text{t} \sigma_z^2}\textbf{P}\boldsymbol{\Sigma}\textbf{V}^\text{H}\textbf{F}_\text{RF}^{\text{opt}}  \textbf{F}_\text{RF}^{{\text{opt}}^\text{H} }\textbf{V} \boldsymbol{\Sigma}^\text{H}  \Big) \\ \nonumber
& =\max_{\text{trace(\textbf{P})}\leq 1}\log_2 \text{det} \Big (\textbf{I}_K+\frac{P_\text{t}}{ \sigma_z^2}\frac{\pi}{4}\textbf{P}\boldsymbol{\Sigma}\textbf{V}^\text{H}\textbf{F}_\text{d}  \textbf{F}_\text{d}^\text{H} \textbf{V} \boldsymbol{\Sigma}^\text{H}  \Big)  \\ \nonumber
&=C_\text{sum}(\frac{\pi}{4}P_\text{t}, \textbf{U}^\text{H} \textbf{H}\textbf{F}_\text{d}) = C_\text{sum}(\frac{\pi}{4} P_\text{t},\frac{1}{\sqrt{\Gamma_\text{t}}}\textbf{H}_\text{e} \textbf{F}_\text{ZFe})\\ \nonumber
&=K\log_2 (1+\frac{\pi P_\text{t}}{4K \Gamma_\text{t}\sigma_z^2}).
\end{align}
Hence, by letting the baseband precoder for the hybrid beamformer with constant modulus constraint as $\textbf{F}^\text{opt}_\text{B} =\textbf{F}_\text{ZFe}$ combined with $\textbf{F}_\text{RF}^{\text{opt}}$ of Lemma 2, the asymptotically optimal hybrid beamformer is achieved. Finally, it could be easily verified that 
\begin{align}
C_\text{sum}(P_\text{t},\textbf{H})-R_\text{sum}(P_\text{t},\frac{1}{\sqrt{N_\text{t}\Gamma_\text{t}}}\textbf{H}\textbf{F}_\text{RF}^{\text{opt}}\textbf{F}_\text{B}^{\text{opt}})=\lim_{N_\text{t} \to \infty} K\log_2\frac{1+\frac{ P_\text{t}}{K \Gamma_\text{t}\sigma_z^2}}{1+\frac{\pi P_\text{t}}{4K \Gamma_\text{t}\sigma_z^2}}=-K\log_2 (\pi/4).
\end{align}

\section{Proof of Lemma 6}
\label{Appendix:ProofLemma6}
According to Appendix \ref{App:lemma2}, the spectral efficiency achieved by the hybrid beamformer depends on $\frac{1}{\sqrt{N_\text{t}}}\textbf{v}_{k}^\text{H} \textbf{f}_{\text{RF},k}$. When the RF beamformer is set based on (\ref{eq:PSselectionBF}),
\begin{align}
\label{eq:eigproblemSwitchoff}
\lim_{N_\text{t} \to\infty}\frac{1}{\sqrt{N_\text{t}}} \textbf{v}_{k}^\text{H} \textbf{f}_{\text{RF},k}  &=\lim_{N_\text{t} \to\infty}\frac{1}{\sqrt{N_\text{t}}}\sum_{n_\text{t}=1}^{N_\text{t}} V_{n_\text{t}k}^\ast {F}_{\text{RF},n_\text{t}k} =\lim_{N_\text{t} \to\infty}\frac{1}{{N_\text{t}}}\sum_{n_\text{t}=1}^{N_\text{t}}\sqrt{N_\text{t}} V_{n_\text{t}k}^\ast {F}_{\text{RF},n_\text{t}k}=\text{E}[ \tilde{V}_{n_\text{t}k}],
\end{align}
where $ \tilde{V}_{n_\text{t}k}$ is defined as
\begin{align}
\tilde{V}_{n_\text{t}k}= 
\begin{cases}
   0, &  \sqrt{{N}_\text{t}}  \vert V_{n_\text{t}k} \vert \leq \alpha ,\\
   \sqrt{N_\text{t}} \vert V_{n_\text{t}k} \vert, &  \alpha < \sqrt{{N}_\text{t}} \vert V_{n_\text{t}k} \vert.
  \end{cases}
  \end{align}
Theorem 1 states that $\sqrt{N_\text{t}}\vert V_{n_\text{t}k}\vert $ follows a Rayleigh distribution with parameter $\sigma_\text{R}$. As a result, the PDF of $ \tilde{V}_{n_\text{t}k}$ is expressed as
\begin{align}
\text{Pr}(\tilde{V})= 
\begin{cases}
   \text{Pr}(\sqrt{{N}_\text{t}} \vert V \vert \leq \alpha) \delta (0), &   \tilde{V} \leq \alpha  ,\\
   \frac{\tilde{V}}{\sigma_\text{R}^2} \text{e}^{-\tilde{V}^2/2\sigma_\text{R}^2},& \alpha <  \tilde{V}.
  \end{cases}
  \end{align} 
The expected value of $\tilde{V}$ is calculated as
\begin{align}
\label{eq:Vtildecalculation1}
\text{E}[\tilde{V}(\alpha)]&= \int_{-\infty}^{+\infty} \tilde{V}\text{Pr}(\tilde{V}) d \tilde{V}=\int_{\alpha}^{+\infty}  \frac{\tilde{V}^2}{\sigma_\text{R}^2} \text{e}^{-\tilde{V}^2/2\sigma_\text{R}^2}d \tilde{V} \\ \nonumber
&= \int_{0}^{+\infty}  \frac{\tilde{V}^2}{\sigma_\text{R}^2} \text{e}^{-\tilde{V}^2/2\sigma_\text{R}^2} d \tilde{V} -\int_{0}^{\alpha}  \frac{\tilde{V}^2}{\sigma_\text{R}^2} \text{e}^{-\tilde{V}^2/2\sigma_\text{R}^2} d \tilde{V} \stackrel{(g)}{=}\sigma_\text{R} \sqrt{\frac{\pi}{2}}-\int_{0}^{\alpha}  \frac{\tilde{V}^2}{\sigma_\text{R}^2} \text{e}^{-\tilde{V}^2/2\sigma_\text{R}^2} d \tilde{V} \\ \nonumber
&\stackrel{(h)}{=}\sigma_\text{R} \sqrt{\frac{\pi}{2}}- \sqrt{2}\sigma_\text{R} \Big( \frac{\sqrt{\pi}}{2}\text{erf}(\frac{\alpha}{\sqrt{2}\sigma_\text{R}}) - \frac{\alpha}{\sqrt{2}\sigma_\text{R}}\text{e}^{-\alpha^2/2\sigma_\text{R}^2} \Big) =\frac{\sqrt{\pi}}{2}+ \alpha \text{e}^{-\alpha^2} -   \frac{\sqrt{\pi}}{2}\text{erf}(\alpha) ,
\end{align}
where $(g)$ and $(h)$ are derived from \cite{gradshteyn2007}. Moreover, the cumulative distribution function (CDF) of $\sqrt{{N}_\text{t}} \vert V \vert \leq \alpha$ is expressed as $\text{Pr}(\vert V_{n_\text{t}k} \vert \leq \alpha) =\beta /100=1-\text{e}^{-\alpha^2/2\sigma_\text{R}^2}$, hence $\alpha=\sqrt{-\text{ln}(1-\beta/100)}$ where $\beta$ is the percentage of the phase shifters that are turned off. It could be easily shown that $\frac{1}{\sqrt{N_\text{t}}}\vert \textbf{v}_{k}^\text{H} \textbf{f}_{\text{RF},k'} \vert=0$, $\forall k \neq k'$ and $\frac{1}{\sqrt{N_\text{t}}} \textbf{F}_\text{d}^\text{H} \textbf{F}_\text{RF}$ becomes a diagonal matrix with equal diagonal elements. Hence, the baseband precoder matrix becomes $\textbf{F}_\text{B}=\textbf{I}_K$. Applying the same phase shifter selection scheme at the receiver side, it can be easily verified that $\Gamma_\text{t}=(1-\beta/100)N_\text{t}$, $\Gamma_\text{r}=(1-\beta/100)N_\text{r}$, $\textbf{R}_\text{n}=1/\Gamma_\text{r} \textbf{W}_\text{B}^\text{H}\textbf{W}_\text{RF}^\text{H}\textbf{W}_\text{RF}\textbf{W}_\text{B}=\textbf{I}_K$ and $1/\sqrt{N_\text{t}} \textbf{F}_\text{d}^\text{H} \textbf{F}_\text{RF} =\text{E}[\tilde{V}(\alpha)] \textbf{I}_K$. Similar to Appendix \ref{App:Lemma3Proof}, the spectral efficiency is expressed as
\begin{align}
\label{eq:RProposedPSSelection}
R_\beta & = \lim_{N_\text{t}\to \infty} \lim_{N_\text{r}\to \infty}  \log_2  \text{det} \Big ( \frac{\rho}{(1-\beta)^2 N_\text{t}N_\text{r}}  \textbf{R}_\text{n}^{-1}\textbf{W}_\text{B}^\text{H}\textbf{W}_\text{RF}^\text{H}\textbf{H} \textbf{F}_\text{RF} \textbf{F}_\text{B} \textbf{P}\textbf{F}_\text{B}^\text{H}\textbf{F}_\text{RF}^\text{H} \textbf{H}^\text{H}\textbf{W}_\text{RF}\textbf{W}_\text{B} \Big ) \\ \nonumber
&= \lim_{N_\text{t}\to \infty} \lim_{N_\text{r}\to \infty}  \log_2  \text{det} \Big ( \frac{\rho}{(1-\beta)^2 N_\text{t} N_\text{r}}  \textbf{W}_\text{RF}^\text{H}\textbf{U} \boldsymbol{\Sigma} \textbf{V}^\text{H} \textbf{F}_\text{RF}   \textbf{P} \textbf{F}_\text{RF}^\text{H}\textbf{V} \boldsymbol{\Sigma} \textbf{U}^\text{H}\textbf{W}_\text{RF} \Big ) \\ \nonumber
&=\sum_{k=1}^K \log_2 \Bigg( \frac{\rho  P_{kk} \sigma_{kk}^2  }{(1-\beta)^2}\Big(\text{E}[\tilde{V}(\alpha)] \Big )^4 \Bigg)\\ \nonumber
&\stackrel{(i)}{=}\sum_{k=1}^K \log_2 ( \rho  P_{kk} \sigma_{kk}^2)+4K\log_2 \Big(\text{E}[\tilde{V}(\alpha)] \Big )-2K\log_2 (1-\beta),
\end{align}
as the first term after $(i)$ is equal to $C$ at high SNR. \hfill\(\Box\)

\section*{Acknowledgement}

The research leading to these results has received funding from the European Union Seventh Framework Programme (FP7/2007-2013) under grant agreement $\text{n}^\circ$619563 (MiWaveS). We would also like to acknowledge the support of the University of Surrey 5GIC members for this work.
\ifCLASSOPTIONcaptionsoff
  \newpage
\fi

\bibliographystyle{IEEEtran}
\bibliography{IEEEabrv,References}

\begin{thebibliography}{10}
\providecommand{\url}[1]{#1}
\csname url@samestyle\endcsname
\providecommand{\newblock}{\relax}
\providecommand{\bibinfo}[2]{#2}
\providecommand{\BIBentrySTDinterwordspacing}{\spaceskip=0pt\relax}
\providecommand{\BIBentryALTinterwordstretchfactor}{4}
\providecommand{\BIBentryALTinterwordspacing}{\spaceskip=\fontdimen2\font plus
\BIBentryALTinterwordstretchfactor\fontdimen3\font minus
  \fontdimen4\font\relax}
\providecommand{\BIBforeignlanguage}[2]{{%
\expandafter\ifx\csname l@#1\endcsname\relax
\typeout{** WARNING: IEEEtran.bst: No hyphenation pattern has been}%
\typeout{** loaded for the language `#1'. Using the pattern for}%
\typeout{** the default language instead.}%
\else
\language=\csname l@#1\endcsname
\fi
#2}}
\providecommand{\BIBdecl}{\relax}
\BIBdecl

\bibitem{MasiveMIMORusek}
F.~Rusek, D.~Persson, B.~K. Lau, E.~Larsson, T.~Marzetta, O.~Edfors, and
  F.~Tufvesson, ``Scaling up {MIMO: O}pportunities and challenges with very
  large arrays,'' \emph{Signal Processing Magazine, IEEE}, vol.~30, no.~1, pp.
  40--60, January 2013.

\bibitem{Arrayprocessing}
L.~Godara, ``Application of antenna arrays to mobile communications. ii.
  beam-forming and direction-of-arrival considerations,'' \emph{Proceedings of
  the IEEE}, vol.~85, no.~8, pp. 1195--1245, August 1997.

\bibitem{AlkhatibHeathMagazine}
A.~Alkhateeb, J.~Mo, N.~Gonzalez-Prelcic, and R.~Heath, ``{MIMO} precoding and
  combining solutions for millimeter-wave systems,'' \emph{Communications
  Magazine, IEEE}, vol.~52, no.~12, pp. 122--131, December 2014.

\bibitem{MolischStatisticsBased2006}
P.~Sudarshan, N.~Mehta, A.~Molisch, and J.~Zhang, ``Channel statistics-based
  {RF} pre-processing with antenna selection,'' \emph{IEEE Transactions on
  Wireless Communications}, vol.~5, no.~12, pp. 3501--3511, December 2006.

\bibitem{JSDM2013}
A.~Adhikary, J.~Nam, J.-Y. Ahn, and G.~Caire, ``Joint spatial division and
  multiplexing - the large-scale array regime,'' \emph{IEEE Transactions on
  Information Theory}, vol.~59, no.~10, pp. 6441--6463, October 2013.

\bibitem{JSDM2014}
A.~Adhikary, E.~Al~Safadi, M.~Samimi, R.~Wang, G.~Caire, T.~Rappaport, and
  A.~Molisch, ``Joint spatial division and multiplexing for mm-{W}ave
  channels,'' \emph{IEEE Journal on Selected Areas in Communications}, vol.~32,
  no.~6, pp. 1239--1255, June 2014.

\bibitem{PhaseOnlyLiu2014}
A.~Liu and V.~Lau, ``Phase only {RF} precoding for massive {MIMO} systems with
  limited {RF} chains,'' \emph{IEEE Transactions on Signal Processing},
  vol.~62, no.~17, pp. 4505--4515, September 2014.

\bibitem{SpatiallySparsePrecodingAyach}
O.~El~Ayach, S.~Rajagopal, S.~Abu-Surra, Z.~Pi, and R.~Heath, ``Spatially
  sparse precoding in millimeter wave {MIMO} systems,'' \emph{IEEE Transactions
  on Wireless Communications}, vol.~13, no.~3, pp. 1499--1513, March 2014.

\bibitem{SohailICC2015}
S.~Payami, M.~Shariat, M.~Ghoraishi, and M.~Dianati, ``Effective {RF} codebook
  design and channel estimation for millimeter wave communication systems,''
  \emph{IEEE International Conference on Communication (ICC) Workshops}, pp.
  1226--1231, June 2015.

\bibitem{NiDL15}
\BIBentryALTinterwordspacing
W.~Ni, X.~Dong, and W.~Lu, ``Near-optimal hybrid processing for massive {MIMO}
  systems via matrix decomposition,'' \emph{CoRR}, vol. abs/1504.03777, 2015.
  [Online]. Available: \url{http://arxiv.org/abs/1504.03777}
\BIBentrySTDinterwordspacing

\bibitem{MOlischAntennaselection2005Journal}
X.~Zhang, A.~Molisch, and S.-Y. Kung, ``Variable-phase-shift-based
  {RF}-baseband codesign for {MIMO} antenna selection,'' \emph{IEEE
  Transactions on Signal Processing}, vol.~53, no.~11, pp. 4091--4103, November
  2005.

\bibitem{HeathMUMIMO}
A.~Alkhateeb, G.~Leus, and R.~Heath, ``Limited feedback hybrid precoding for
  multi-user millimeter wave systems,'' \emph{IEEE Transactions on Wireless
  Communications}, vol.~PP, no.~99, pp. 1--1, 2015.

\bibitem{PerAntennaPi2012}
Z.~Pi, ``Optimal transmitter beamforming with per-antenna power constraints,''
  \emph{IEEE International Conference on Communications (ICC)}, pp. 3779--3784,
  June 2012.

\bibitem{FoadTorontoApril2015}
S.~Sohrabi, , and Y.~Wei, ``Hybrid digital and analog beamforming design for
  large-scale {MIMO} systems,'' \emph{IEEE International Conference on
  Acoustics, Speech and Signal Processing (ICASSP)}, April 2015.

\bibitem{FoadTorontoApril2015Digital}
------, ``Hybrid beamforming with finite-resolution phase shifters for
  large-scale {MIMO} systems,'' \emph{IEEE International Workshop on Signal
  Processing for Wireless Communications (SPAWC)}, July 2015.

\bibitem{Tse2005Book}
D.~Tse and P.~Viswanath, \emph{Fundamentals of Wireless Communication}.\hskip
  1em plus 0.5em minus 0.4em\relax New York, NY, USA: Cambridge University
  Press, 2005.

\bibitem{Goldsmith2003}
A.~Goldsmith, S.~Jafar, N.~Jindal, and S.~Vishwanath, ``Capacity limits of
  {MIMO} channels,'' \emph{IEEE Journal on Selected Areas in Communications},
  vol.~21, no.~5, pp. 684--702, June 2003.

\bibitem{Cover:2006:EIT:1146355}
T.~M. Cover and J.~A. Thomas, \emph{Elements of Information Theory (Wiley
  Series in Telecommunications and Signal Processing)}.\hskip 1em plus 0.5em
  minus 0.4em\relax Wiley-Interscience, 2006.

\bibitem{GrassmanianBF}
D.~Love, R.~Heath, and T.~Strohmer, ``Grassmannian beamforming for
  multiple-input multiple-output wireless systems,'' \emph{IEEE International
  Conference on Communications (ICC)}, vol.~4, pp. 2618--2622 vol.4, May 2003.

\bibitem{ECP1294}
M.~Spruill, ``Asymptotic distribution of coordinates on high dimensional
  spheres,'' \emph{Electron. Commun. Probab.}, vol.~12, pp. no. 23, 234--247,
  2007.

\bibitem{citeulike:9070763}
C.~Walck, \emph{{Hand-book on statistical distributions for experimentalists}},
  December 1996.

\bibitem{BeamsteeringAyach2012}
O.~El~Ayach, R.~Heath, S.~Abu-Surra, S.~Rajagopal, and Z.~Pi, ``The capacity
  optimality of beam steering in large millimeter wave {MIMO} systems,''
  \emph{IEEE 13th International Workshop on Signal Processing Advances in
  Wireless Communications (SPAWC)}, pp. 100--104, June 2012.

\bibitem{Brand200620}
M.~Brand, ``Fast low-rank modifications of the thin singular value
  decomposition,'' \emph{Linear Algebra and its Applications}, vol. 415, no.~1,
  pp. 20--30, 2006.

\bibitem{WishvanathTse}
P.~Viswanath and D.~N.~C. Tse, ``Sum capacity of the vector gaussian broadcast
  channel and uplink-downlink duality,'' \emph{IEEE Transactions on Information
  Theory}, vol.~49, no.~8, pp. 1912--1921, August 2003.

\bibitem{PhaseShifterSurvey}
Y.~Zheng and C.~E. Saavedra, ``An ultra-compact {CMOS} variable phase shifter
  for 2.4-{GHz} {ISM} applications,'' \emph{IEEE Transactions on Microwave
  Theory and Techniques}, vol.~56, no.~6, pp. 1349--1354, June 2008.

\bibitem{SwitchesSurvey}
N.~A. Talwalkar, C.~P. Yue, H.~Gan, and S.~S. Wong, ``Integrated {CMOS}
  transmit-receive switch using {LC}-tuned substrate bias for 2.4-{GH}z and
  5.2-{GH}z applications,'' \emph{IEEE Journal of Solid-State Circuits},
  vol.~39, no.~6, pp. 863--870, June 2004.

\bibitem{TulinoMatrix}
A.~M. Tulino and S.~Verd\'{u}, ``Random matrix theory and wireless
  communications,'' \emph{Now Publishers Inc.}, 2004.

\bibitem{1207369}
G.~Caire and S.~Shamai, ``On the achievable throughput of a multiantenna
  {G}aussian broadcast channel,'' \emph{IEEE Transactions on Information
  Theory}, vol.~49, no.~7, pp. 1691--1706, July 2003.

\bibitem{HighSNRjindal}
N.~Jindal, ``High {SNR} analysis of {MIMO} broadcast channels,''
  \emph{Proceedings of International Symposium on Information Theory (ISIT)},
  pp. 2310--2314, September 2005.

\bibitem{gradshteyn2007}
I.~S. Gradshteyn and I.~M. Ryzhik, ``Table of integrals, series, and
  products,'' \emph{Elsevier Academic Press, Amsterdam}, 2007.

\end{thebibliography}

\end{document}